\documentclass[useAMS,usenatbib]{mn2e}
\pdfoutput=1
\usepackage{aas_macros}
\usepackage{graphicx}
\usepackage{dcolumn}
\usepackage{amssymb,amsmath,bm}
\usepackage{color}
\usepackage[colorlinks,linkcolor=red,citecolor=blue,urlcolor=blue ]{hyperref}
\usepackage{multirow}
\usepackage{enumitem}
\usepackage[utf8]{inputenc}
\usepackage{lipsum}

\newcommand{\lotss}{LoTSS}
\newcommand{\nv}{\hat{\bf n}}
\newcommand{\mJy}{\,{\rm mJy}}
\newcommand{\planck}{{\sl Planck}}

\title[Cross-correlating radio continuum surveys and CMB lensing]{Cross-correlating radio continuum surveys and CMB lensing: constraining redshift distributions, galaxy bias and cosmology}
\author[Alonso et al.]{David Alonso$^1$, Emilio Bellini$^1$, Catherine Hale$^2$, Matt J.~Jarvis$^{1,3}$,\newauthor Dominik J.~Schwarz$^{4}$ \\
$^{1}$Astrophysics, University of Oxford, DWB, Keble Road, Oxford OX1 3RH, UK\\
$^{2}$CSIRO Astronomy and Space Science, PO Box 1130, Bentley WA 6102, Australia\\
$^{3}$Department of Physics, University of the Western Cape, Bellville 7535, South Africa \\
$^{4}$Fakult\"at f\"ur Physik, Universit\"at Bielefeld, Postfach 100131, 33501 Bielefeld, Germany}
\begin{document}
  \date{\today}
  \pagerange{1--19} \pubyear{2020}
  \maketitle

  \begin{abstract}
    We measure the harmonic-space auto-power spectrum of the galaxy overdensity in the LOFAR Two-metre Sky Survey (LoTSS) First Data Release and its cross correlation with the map of the lensing convergence of the cosmic microwave background (CMB) from the \planck{} collaboration. We report a $\sim5\sigma$ detection of the cross-correlation. We show that the combination of the clustering power spectrum and CMB lensing cross-correlation allows us to place constraints on the high-redshift tail of the redshift distribution, one of the largest sources of uncertainty in the use of continuum surveys for cosmology. Our analysis shows a preference for a broader redshift tail than that predicted by the photometric redshifts contained in the LoTSS value added catalog, as expected, and more compatible with predictions from simulations and spectroscopic data. Although the ability of CMB lensing to constrain the width and tail of the redshift distribution could also be valuable for the analysis of current and future photometric weak lensing surveys, we show that its performance relies strongly on the redshift evolution of the galaxy bias. Assuming the redshift distribution predicted by the Square Kilometre Array Design simulations, we use our measurements to place constraints on the linear bias of radio galaxies and the amplitude of matter inhomogeneities $\sigma_8$, finding $\sigma_8=0.69^{+0.14}_{-0.21}$ assuming the galaxy bias scales with the inverse of the linear growth factor, and $\sigma_8=0.79^{+0.17}_{-0.32}$ assuming a constant bias.
  \end{abstract}


  \section{Introduction}\label{sec:intro}
    Radio continuum surveys have long attracted the attention of the cosmological community as deep tracers of the large-scale structure, able to probe enormous swathes of the Universe. Several forecasting studies have been carried out to explore the cosmological potential of these probes with the Square Kilometre Array (SKA) \citep{2012MNRAS.424..801R,2015aska.confE..18J}. In particular, it has been argued that, using different radio galaxy populations as individual tracers, continuum surveys could be useful in detecting ultra-large scale effects, such as relativistic lightcone effects and the impact of primordial non-Gaussianity \citep{2014MNRAS.442.2511F,2015PhRvD..92f3525A,2019JCAP...09..025B,Gomes2020}.
      
    Extragalactic radio continuum surveys produce a very different view of the Universe compared to those at shorter wavelengths (e.g. in the optical and IR). At $\sim$0.1-10 GHz, the dominant mechanism for radio continuum emission is from synchrotron radiation \citep{Condon1992}, which originates from relativistic electrons spiralling in the magnetic fields. Due to this, radio observations are typically dominated by two main galaxy populations: Star Forming Galaxies (SFGs) and Active Galactic Nuclei (AGN). SFGs are often classed into normal starforming galaxies (with star formation rates, SFR$\lesssim 100$\,M$_{\odot}$yr$^{-1}$) as well as starburst galaxies, where intense periods of star formation are present (SFR$\gtrsim 100$\,M$_{\odot}$yr$^{-1}$). For AGN, there is a diverse polychotomy of sources as well as classification schemes. This includes classifying sources based on orientation to the observer \citep{Antonucci1993,Urry1995}, morphology \citep[e.g. Faranoff-Riley Type I and II sources][]{Fanaroff1974}, or the accretion mechanism of material onto the central supermassive black hole \cite[High/Low excitation radio galaxies, e.g.][]{Best2012, Heckman2014}. A key advantage in observing both AGN and SFGs at these frequencies is that  attenuation of emission by dust, in the inter-galactic medium and along the line of sight, is negligible. This is especially useful for the study of SFGs, where the radio emission can be used as an unbiased measurement of SFR \citep[see e.g.][]{Bell2003,Davies2017,Gurkan2018}.
  
    Moreover, at these low frequencies, the instantaneous field of view from radio surveys is often large. For example, the LOw Frequency ARray \citep[LOFAR;][]{vanHaarlem2013} has a field of view of $\sim$30 sq. deg at 150 MHz, the Meer Karoo Array Telescope \citep[MeerKAT;][]{Jonas2009} has a field of view of $\sim$1 sq. deg at 1.2 GHz and the Australian Square Kilometre Array Pathfinder \citep[ASKAP;][]{Johnston2007} has an instantaneous field of view of $\sim$30 sq deg, due to its use of phased array feeds (PAFs). These large fields of view for radio interferometers are crucial for producing large, contiguous observations of the celestial sphere, which is especially advantageous for studying large-scale structure at large angular separations. As such, there exist many large sky surveys at radio frequencies, including: the NRAO VLA Sky Survey \citep[NVSS;][at 1.4 GHz]{Condon1998}, TIFR GMRT Sky Survey \citep[TGSS-ADR;][at 150 MHz]{Intema2017}, Sydney University Molongolo Sky Survey \citep[SUMSS;][]{Mauch2003}. Further surveys such as the Evolutionary Map of the Universe \citep[EMU;][]{Norris2011} and the LOFAR Two-metre Sky Survey \citep[LoTSS][]{2019A&A...622A...1S}, will soon provide a huge leap by combining both depth and sky coverage \citep[e.g. see Fig. 1 of ][]{2019A&A...622A...1S}.

    Already though, the first data release of LOFAR Two-metre Sky Survey \citep{2019A&A...622A...1S} has generated a low frequency radio continuum survey, covering 424 sq. deg at 144 MHz to a depth of $\sim$70 $\mu$Jy/beam. This is one of the deepest surveys available at this frequency that covers a large area of sky, ideal for studies of large scale structure \citep[see][]{2019arXiv190810309S}.

    However, the main complication in using continuum data for cosmology is the absence of reliable redshifts for a large fraction of sources. This leads to large uncertainties in the redshift distribution of different samples, and in general on the redshift evolution of any of their properties, such as the galaxy bias or host halo masses. The large-scale structure analysis of continuum samples with optical matches, has been a useful method to address these issues and improve our understanding of the clustering properties of radio galaxies \citep{Lindsay2014a,Hale:2017wub,2019arXiv190810309S}.

    Further information can be gained through cross-correlations with other tracers of the large-scale structure \citep[e.g.][]{Lindsay2014b}. In particular, since the gravitational lensing of the Cosmic Microwave Background (CMB) is sensitive to the distribution of matter inhomogeneities to very high redshifts, it constitutes an ideal probe to cross-correlate with deep radio continuum data \citep[e.g.][]{2014A&A...571A..17P,2015MNRAS.451..849A}, including in the context of de-lensing \citep{2016PhRvD..93d3527N}. In this paper, we will explore the combination of the auto-correlation of the LoTSS survey and its cross-correlation with CMB lensing data from the \planck{} satellite. In particular, we will show how the different dependence of both correlations on the galaxy bias and redshift distribution of the sample allows us to use this combination to simultaneously constrain the bias and the high-redshift tail of the redshift distribution. This possibility is also relevant in the context of weak lensing analyses with current and future deep imaging surveys \citep{2020A&A...640L..14W,2020arXiv200715635H}, for which the robustness of their cosmological constraints will rely heavily on their ability to calibrate the redshift distributions of their samples.

    This paper is structured as follows: in Section \ref{sec:theory} we present the theoretical background describing the auto- and cross-correlation signals. Section \ref{sec:data} presents the different datasets used in our analysis. The methods used to extract the auto- and cross-correlations, and to analyse them, are described in Section \ref{sec:methods}. The main results are then presented and discussed in Section \ref{sec:results}, and we conclude in Section \ref{sec:conc}.

  \section{Theory}\label{sec:theory}
    \subsection{Galaxy overdensity and CMB lensing}\label{ssec:theory.fields}
      The two probes studied here are the projected overdensity of galaxies in the LoTSS sample, $\delta_g$, and the gravitational lensing convergence of the CMB, $\kappa$.
      
      The projected overdensity quantifies the over-abundance of galaxies in a given sky position $\nv$, $N_g(\nv)$ with respect to the sky average $\bar{N}_g$
      \begin{equation}
        \delta_g(\nv)\equiv\frac{N_g(\nv)-\bar{N}_g}{\bar{N}_g},
      \end{equation}
      and is related to the three-dimensional galaxy overdensity, $\Delta_g({\bf x},z)$ (defined in a similar way as the fluctuation in the comoving number density of galaxies at redshift $z$) through
      \begin{equation}\label{eq:delta_g}
        \delta_g(\nv)=\int dz\,\frac{dp}{dz}\,\Delta_g(\chi(z)\nv,z),
      \end{equation}
      where $dp/dz$ is the redshift distribution of the galaxy sample normalized to 1 when integrated over $z$, and $\chi$ is the comoving radial distance.

      The lensing convergence $\kappa(\nv)$ quantifies the distortion in the trajectories of the CMB photons caused by the gravitational potential of the intervening matter structures, and is defined to be proportional to the divergence of the deflection in the photon arrival angle $\pmb{\alpha}$: $\kappa\equiv-\nabla\cdot\pmb{\alpha}/2$. As such, $\kappa$ is an unbiased tracer of the matter density fluctuations $\Delta_m({\bf x},z)$, and is related to them through:
      \begin{equation}
        \kappa(\nv)=\int_0^{\chi_{\rm LSS}} d\chi\,\frac{3H_0^2\Omega_m}{2a}\chi\frac{\chi_{\rm LSS}-\chi}{\chi_{\rm LSS}}\Delta_m(\chi\nv,z(\chi))
      \end{equation}
      where $H_0$ is the Hubble constant, $\Omega_m$ is the fractional matter density, $a=1/(1+z)$ is the scale factor, and $\chi_{\rm LSS}$ is the comoving distance to the surface of last scattering.
      
      We will use the harmonic-space correlation between $\kappa$ and $\delta_g$ to study the connection between $\Delta_g$ and $\Delta_m$.

    \subsection{Angular power spectra}\label{ssec:theory.cls}
      Both $\kappa$ and $\delta_g$ can be generically described as a 2-dimensional field $u(\nv)$ related to an underlying 3-dimensional quantity $U({\bf x},z)$ through a projection onto the two-dimensional sphere with a given radial kernel $W_u(\chi)$:
      \begin{equation}
        u(\nv)=\int d\chi\,W_u(\chi)\,U(\chi\nv,z(\chi)).
      \end{equation}

      Any such projected quantity can be decomposed in terms of its spherical harmonic coefficients $u_{\ell m}$, the covariance of which is the so-called angular power spectrum:
      \begin{equation}
        \langle u_{\ell m}v^*_{\ell'm'}\rangle=\delta_{\ell\ell'}\delta_{mm'}C_\ell^{uv}
      \end{equation}
      The angular power spectrum can be related to the power spectrum of the 3D fields $P_{UV}$ through
      \begin{equation}\label{eq:cl_limber}
        C^{uv}_\ell = \int \frac{d\chi}{\chi^2}W_u(\chi)W_v(\chi)P_{UV}\left(k=\frac{\ell+1/2}{\chi},z(\chi)\right),
      \end{equation}
      where $P_{UV}(k,z)$ is the variance of the Fourier coefficients of $U$ and $V$:
      \begin{equation}
        \langle U({\bf k},z)V^*({\bf k}',z)\rangle\equiv(2\pi)^3\delta({\bf k}-{\bf k}')P_{UV}(k,z).
      \end{equation}
      We use the following definition for the Fourier transform:
      \begin{equation}
        U({\bf x},z)=\int\frac{dk^3}{(2\pi)^2}e^{i{\bf k}\cdot{\bf x}}U({\bf k},z).
      \end{equation}
      
      For the two fields under consideration, the radial kernels are given by
      \begin{align}
        &W_g(\chi)=\frac{H(z)}{c}\frac{dp}{dz},\\
        &W_\kappa(\chi)=f_\ell\frac{3H_0^2\Omega_m}{2a}\chi\frac{\chi_{\rm LSS}-\chi}{\chi_{\rm LSS}}\,\Theta(\chi_{\rm LSS}-\chi)
      \end{align}
      where $H(z)$ is the expansion rate, and $\Theta(x)$ is the Heaviside function. Note that we have included a scale-dependent prefactor $f_\ell$ in the lensing kernel, given by
      \begin{equation}
        f_\ell=\frac{\ell(\ell+1)}{(\ell+1/2)^2}\simeq1.
      \end{equation}
      The presence of this factor is only relevant for $\ell\lesssim10$, and accounts for the fact that $\kappa$ is related to $\Delta_m$ through the angular Laplacian of the gravitational potential $\Phi$. We have also neglected the effects of lensing magnification, which are much smaller than our statistical uncertainties \citep{2015MNRAS.451..849A}\footnote{This may not be the case for the final LoTSS sample with smaller uncertainties.}.
      
      It is worth noting that Eq. \ref{eq:cl_limber} is stictly valid only in the Limber approximation, where the spherical Bessel functions are approximated by
      \begin{equation}
        j_\ell(x)\simeq\sqrt{\frac{\pi}{2\ell+1}}\delta(x-\ell-1/2).
      \end{equation}
      This is accurate when the radial kernels $W_u$ are broader than the typical correlation length of the density inhomogeneities, as is the case for both $\delta_g$ and $\kappa$.

      The final ingredient of the model needed to describe the measured angular power spectra is the auto- and cross-spectra between the matter and galaxy overdensity fields, $P_{gg}(k,z)$, $P_{gm}(k,z)$ and $P_{mm}(k,z)$. On the scales $k\lesssim 0.2\,h\,{\rm Mpc}^{-1}$ we probe here, we assume a linear, scale-independent bias model between both fields: $\Delta_g({\bf k},z)=b_g(z)\Delta_\kappa({\bf k},z)$, such that
      \begin{align}
        &P_{gm}(k,z)=b_g(z)\,P_{mm}(k,z),\\
        &P_{gg}(k,z)=b_g^2(z)\,P_{mm}(k,z).
      \end{align}
      We explore two different models for the redshift dependence of the bias:
      \begin{itemize}
          \item A {\sl constant bias} model, in which $b_g$ does not evolve, and the clustering of galaxies follows the same growth as a function of time as the matter inhomogeneities.
          \item A {\sl constant amplitude} model, in which $b_g$ evolves inversely with the linear growth factor $D(z)$
          \begin{equation}\label{eq:b_grth}
            b_g(z)=b_g/D(z),
          \end{equation}
          where the growth factor $D(z)\equiv\Delta_m({\bf k},z)/\Delta_m({\bf k},z=0)$ describes the evolution of the matter fluctuations on linear scales ($k\rightarrow0$). In this scenario, the amplitude of $\Delta_g$ does not vary as a function of time on linear scales. This would correspond to a galaxy distribution that is fixed at some early time and preserves its large-scale properties unchanged.
      \end{itemize}
      These two models represent the two extremes of the expected redshift evolution of $b_g(z)$. Flux-density limited samples such as the one studied here are expected to have $b_g$ grow with $z$, since more distant sources are intrinsically brighter and are therefore likely to reside in more massive haloes. At the same time, the LoTSS sample contains a wide variety of galaxy types, with different biases and different relative abundances as a function of redshift, and therefore the growth of $b_g(z)$ could be slower than the $1/D(z)$ model.

      Finally, we model the matter power spectrum $P_{mm}(k,z)$ using the {\tt HALOFIT} parameterization of \citet{2012ApJ...761..152T} as implemented in the {\tt CAMB} Boltzmann code \citep{Lewis:1999bs}. We use the Core Cosmology Library \citep[CCL;][]{2019ApJS..242....2C} for all theoretical calculations.

  \section{Data}\label{sec:data}
    \subsection{The LOFAR Two-metre Sky Survey data release 1}\label{ssec:data.lotss}
      The radio continuum sample used here is contained in the first data release (DR1) of the LOFAR Two-metre Sky Survey (LoTSS). Whilst the full LoTSS survey will eventually cover the entire northern sky, this first data release covers a contiguous patch of 424 square degrees in the northern hemisphere over the HETDEX spring field, and contains a total of 325,694 sources. The DR1 is described in detail in \citet{2019A&A...622A...1S}, and the clustering properties of the sample were studied for the first time in \citet{2019arXiv190810309S}. 

      DR1 consists of 58 individual pointings, each covering an area approximately the shape of a circle of radius 1.7$^\circ$ at 6 arcsecond resolution. Each of the 58 pointings was observed for a duration of 8 hours, resulting in an rms noise of $\sim$70 $\mu$Jy/beam. \citet{2019arXiv190810309S} found five of these pointings to result in a poor completeness (at the level of $\sim90\%$ at flux $S\simeq 1\mJy$), we discard these pointings in our analysis as well. The average point-source completeness of the remaining area is approximately $99\%$ level for $S>0.8\mJy$.
      
      Through a combination of likelihood-ratio matching and visual identification, a value added catalogue (VAC) was provided as part of the DR1, which we use for our analysis \citep{2019A&A...622A...2W}. The VAC contains 318,520 sources, of which 73\% have additional optical and or infrared information, primarily from cross-matches with Pan-STARRS and/or WISE. For approximately half of the sample, redshifts were also assigned to sources within the VAC \citep{2019A&A...622A...3D}. The majority of these are photometric redshifts, whilst a minority of sources have associated spectroscopic redshifts. Of those sources included in the VAC, we use a fiducial flux limited sample defined by a cut in total flux of $S>2\mJy$. We additionally discard all sources with a signal-to-noise ratio $S/N<5$. The remaining sample contains 57,928 sources.

    \subsection{The Planck lensing maps}\label{ssec:data.planck}
      We use the publicly available CMB lensing convergence map provided by the \planck{} Collaboration \citep{2018arXiv180706210P}. The data are provided in the form of spherical harmonic coefficients $\kappa_{\ell m}$, which we transform into a HEALPix map \citep{2005ApJ...622..759G} with resolution parameter $N_{\rm side}=2048$, corresponding to an angular size of $\sim2'$. We verified that this choice of resolution has a negligible impact on our results by re-calculating the $\delta_g$-$\kappa$ cross-correlation presented in Section \ref{sec:results} for $N_{\rm side}=256$. When estimating power spectra, we mask this convergence map using the mask provided by Planck, covering around 67\% of the sky. This mask was designed to remove regions dominated by galactic foregrounds as well as strong Sunyaev-Zel'dovich sources. This mask removes around $\sim2\%$ of the LoTSS footprint.

    \subsection{Redshift distributions}\label{ssec:data.nz}
      \begin{figure}
        \centering
        \includegraphics[width=0.49\textwidth]{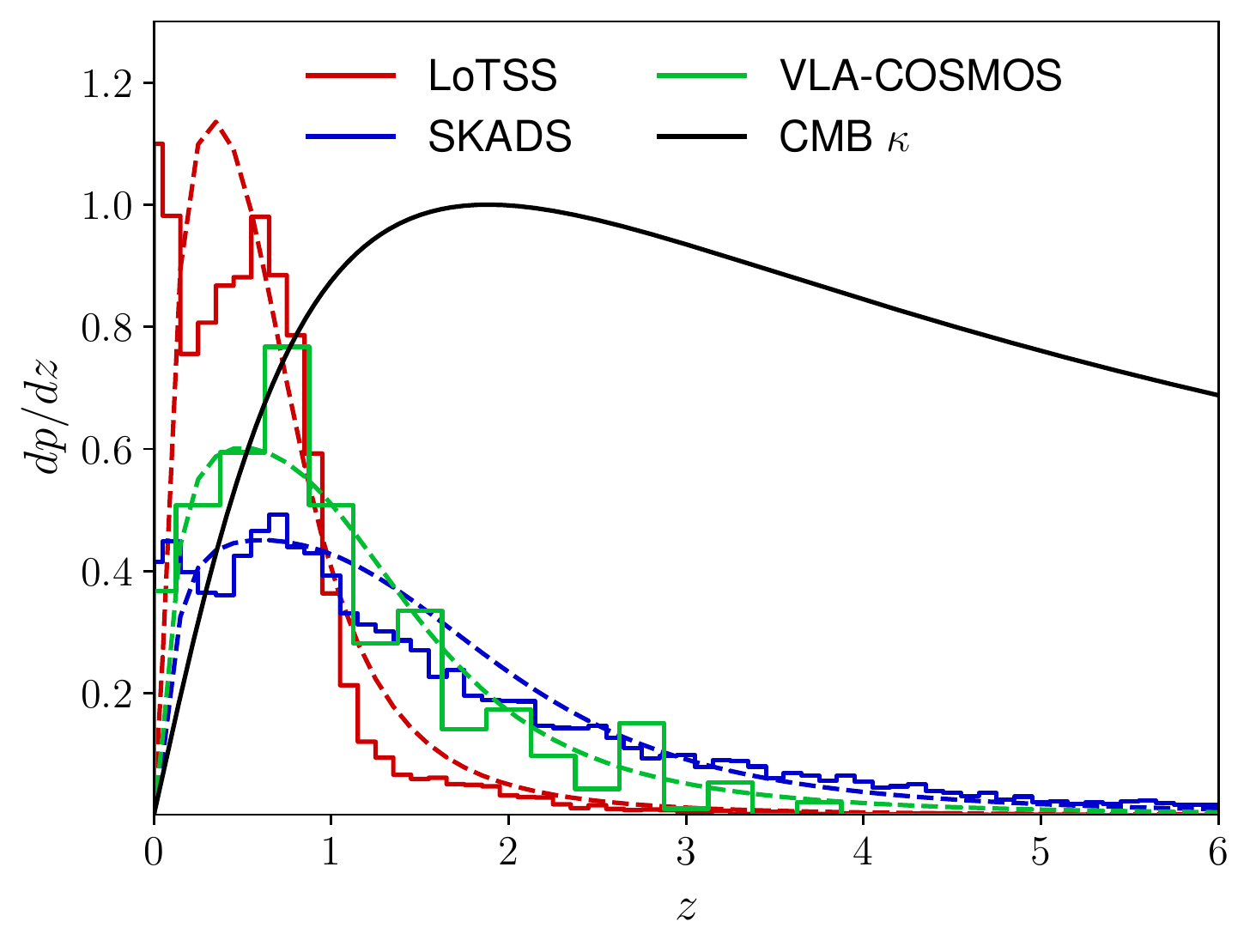}
        \caption{Estimated redshift distribution of the flux-limited sample used in this work. The red histogram shows the distribution estimated from the photometric redshifts contained in the LoTSS value-added catalog. The green histogram shows the estimate from the spectroscopic VLA-COSMOS sample at $3\,{\rm GHz}$ extrapolated to the LOFAR band. The blue histogram shows the distribution obtained from the SKADS simulation, based on existing measurements of the radio luminosity function. The dashed lines are empirical fits to the three distributions using the expression in Eq. \ref{eq:nz_ana} with a single free parameter $z_{\rm tail}$. The black solid line shows the CMB lensing kernel normalized to make it viewable on the same scale. The distributions shown here are plotted as probability distributions $dp/dz$.}\label{fig:nz}
      \end{figure}
      The redshift distribution of the LoTSS sample used here is a key ingredient in order to obtain a theoretical prediction for the clustering auto-spectrum and the CMB lensing cross-spectrum ($dp/dz$ in Eq \ref{eq:delta_g}).
      
      The LoTSS value-added catalog provides photometric redshift (photo-$z$) information for about 51\% of the total sample (and $\sim48\%$ of our flux limited sample), which can be used to estimate this redshift distribution. There are however significant uncertainties associated with these photometric redshifts. Of particular concern is the level to which the subset of the catalog with measured redshifts is representative of the full sample, especially at high redshifts. Although the distribution of radio fluxes in the two samples (the full sample and the subset with measured redshifts), shows no significant differences, this is no guarantee that other selection effects (e.g. cuts in the optical photometry) do not induce significant differences in their redshift distributions. This may be especially important in the tail of high redshift AGN which is typically found in radio surveys. It is likely that the $N(z)$ distribution from \citet{2019A&A...622A...3D} may underestimate the number of high-redshift sources, especially given radio emission does not suffer dust attenuation whilst the cross-matched catalogues is affected by this, especially for distant objects. Another potential source of error would be mis-identification of high-redshift AGNs as low-redshift SFGs. Also, as noted in \citet{2019arXiv190810309S}, the low- and high-redshift differential source count distributions show markedly different shapes. Therefore other estimates of $dp/dz$ are needed in order to assess the reliability of our results. The redshift distribution estimated from the LoTSS VAC is shown in red in Fig. \ref{fig:nz} as a solid line.

      We obtain a second estimate of the redshift distribution from the Very Large Array Cosmic Evolution Survey (VLA-COSMOS) 3 GHz catalog \citep{2017A&A...602A...1S}, which comprises $\sim10,000$ radio sources in a 2 square-degree patch covering the COSMOS field in combination with optical and infrared data. Due to the vast wealth of deep multi-wavelength data in the COSMOS field, 8995 of these sources have measured redshifts. This is likely to therefore provide a much better estimate of the high redshift tail of radio sources. To obtain an estimate of the redshift distribution of the LoTSS sample, we extrapolate the radio fluxes of these objects to the LOFAR 144 MHz band assuming a spectral index $\alpha=-0.7$\footnote{$\alpha$ is defined as $S_{\nu} \propto \nu^\alpha$} and impose the same flux cut of used for our sample (see Section \ref{ssec:data.lotss}). Due to this relatively bright flux cut, the resulting sample contains only 378 galaxies. Although the small size of this sample prevents us from obtaining a well-resolved measurement of the redshift distribution, it is enough to quantify the amplitude of the high-redshift tail in comparison with the LoTSS VAC measurement. The resulting redshift distribution, estimated by binning these galaxies into 20 linear redshift bins in the range $0<z<5$, is shown in green in Fig. \ref{fig:nz}. Although the VLA-COSMOS catalog potentially underestimates the nearest extended objects due to its baseline sensitivity to extended emission, it displays a significantly larger high-redshift tail than the LoTSS VAC. As we will see, this high-redshift tail plays a significant role in determining the relative amplitudes of $C_\ell^{gg}$ and $C_\ell^{g\kappa}$.
      
      Finally, we use the Square Kilometre Array Design Study Simulated Skies (SKADS), in particular the semi-empirical simulation \citep{2008MNRAS.388.1335W}, to obtain a third estimate of the redshift distribution of our sample. This simulation contains a realistic distribution of radio galaxies, including AGN and star-forming galaxies, down to flux densities of $10\,{\rm nJy}$ based on estimates of the radio luminosity function for different source types, on a square 100 deg$^2$ patch. Each source has measured flux  densities at several frequencies in the range 151 MHz to 18 GHz, which we use to infer their flux in the LOFAR band. The redshift distribution is then computed by binning all sources in the simulation with fluxes above our cut of $2\,{\rm mJy}$. The resulting distribution is shown in blue in Fig. \ref{fig:nz}. Much like the distribution inferred from the VLA-COSMOS catalog, the SKADS estimate displays a long high-redshift tail when compared with the LoTSS $dp/dz$. There is evidence though, that SKADS may underestimate the number of SFGs below flux densities of $S_{1.4 GHz}<0.1$ mJy \citep{2017A&A...602A...1S}, however as this is below the survey limits considered in this work, the effect of this is likely to be minimal. Estimates of $dp/dz$ will be improved in the future with deeper observations over well-studied multi-wavelength fields, such as through the MIGHTEE \citep{Jarvis2016} or deep-field LOFAR observations.
      
      This large uncertainty in the true redshift distribution for radio continuum surveys motivates the use of the cross-correlation with CMB lensing as a means to mitigate it. To aid in this study, we fit these three redshift distributions to the same analytic expression
      \begin{equation}\label{eq:nz_ana}
        \frac{dp}{dz}\propto\frac{(z/z_0)^2}{1+(z/z_0)^2}\frac{1}{1+(z/z_{\rm tail})^\gamma}.
      \end{equation}
      The normalising prefactor is determined by the requirement that the integral over $dp/dz$ be unity. We find that $z_0=0.1$ and $\gamma=3.5$ provide a good visual fit to the three redshift distributions. The free variable $z_{\rm tail}$ parametrizes the extent of the high-redshift tail of the distribution, and therefore its width. We find that this expression is able to provide a rough fit to the three redshift distributions with $z_{\rm tail}=0.8,\,1.5$ and $2.0$ for the LoTSS, VLA-COSMOS and SKADS $N(z)$s respectively. These analytic fits are shown as dashed curves in Fig. \ref{fig:nz}. The figure also shows, as a black solid line, the shape of the CMB lensing kernel, which peaks at $z\sim2$ but extends to significantly higher redshifts.

  \section{Methods}\label{sec:methods}
    \subsection{Mean density, sky mask and systematics}\label{ssec:methods.weights}
      \begin{figure*}
        \centering
        \includegraphics[width=0.47\textwidth]{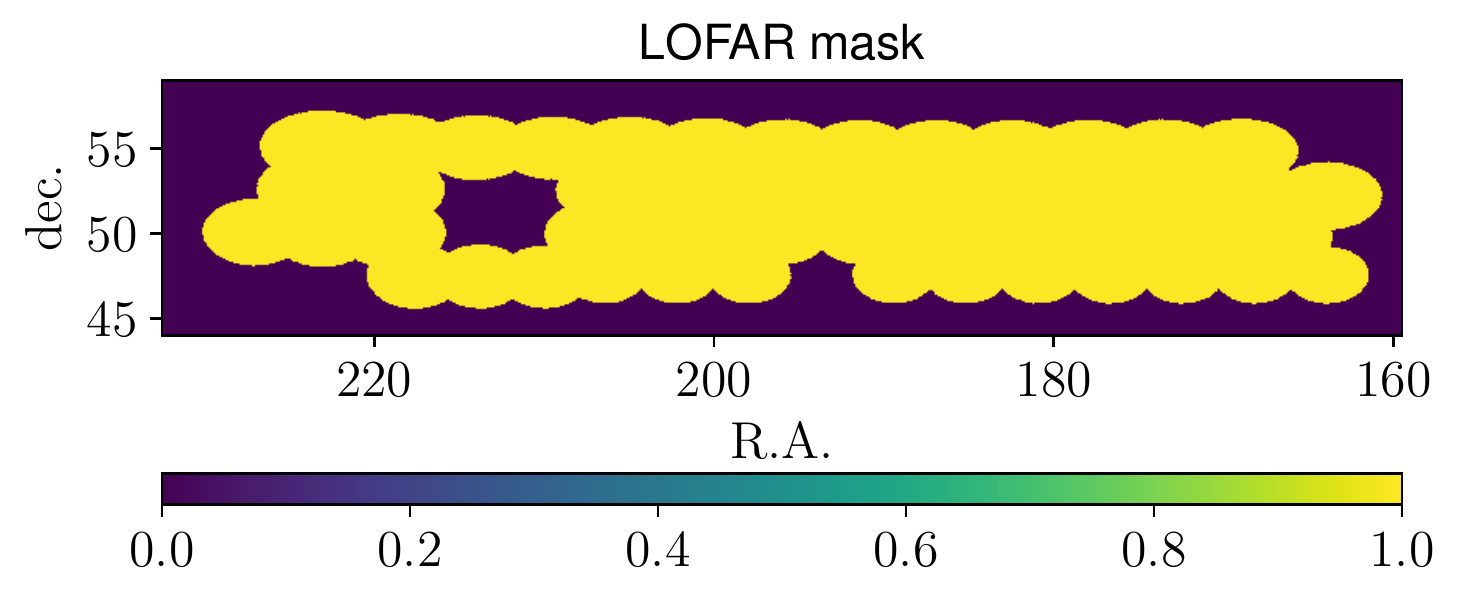}
        \includegraphics[width=0.47\textwidth]{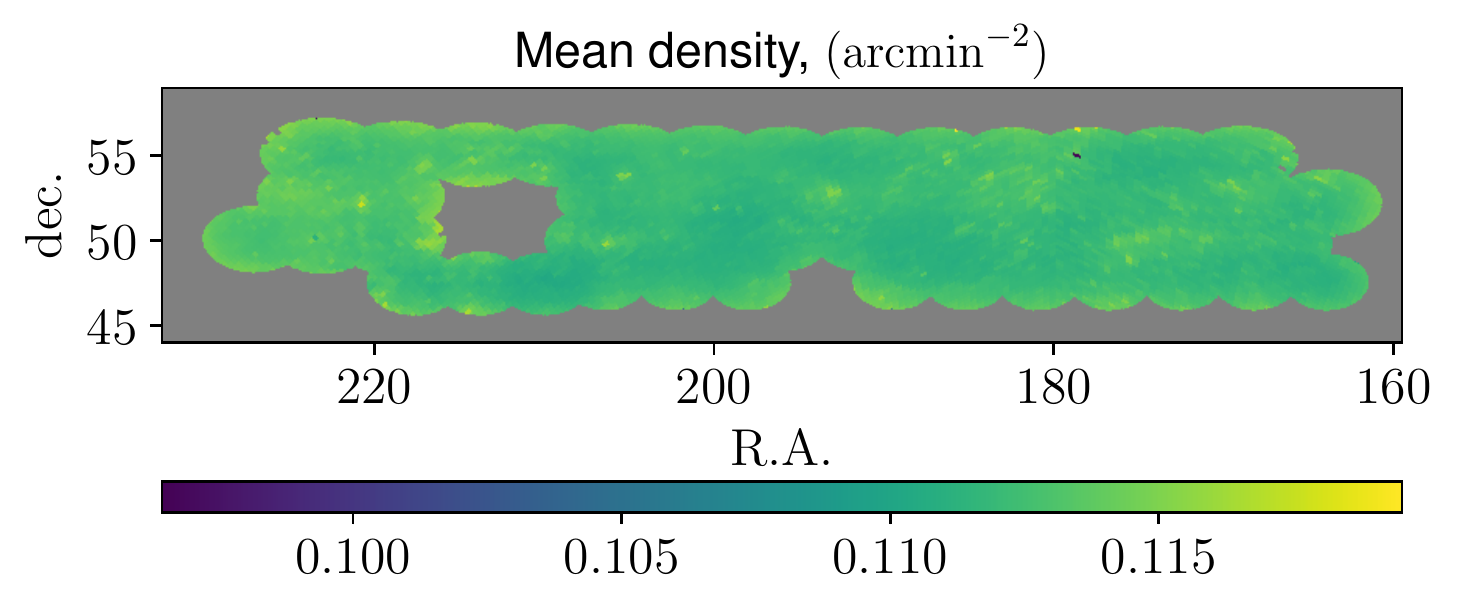}
        \includegraphics[width=0.47\textwidth]{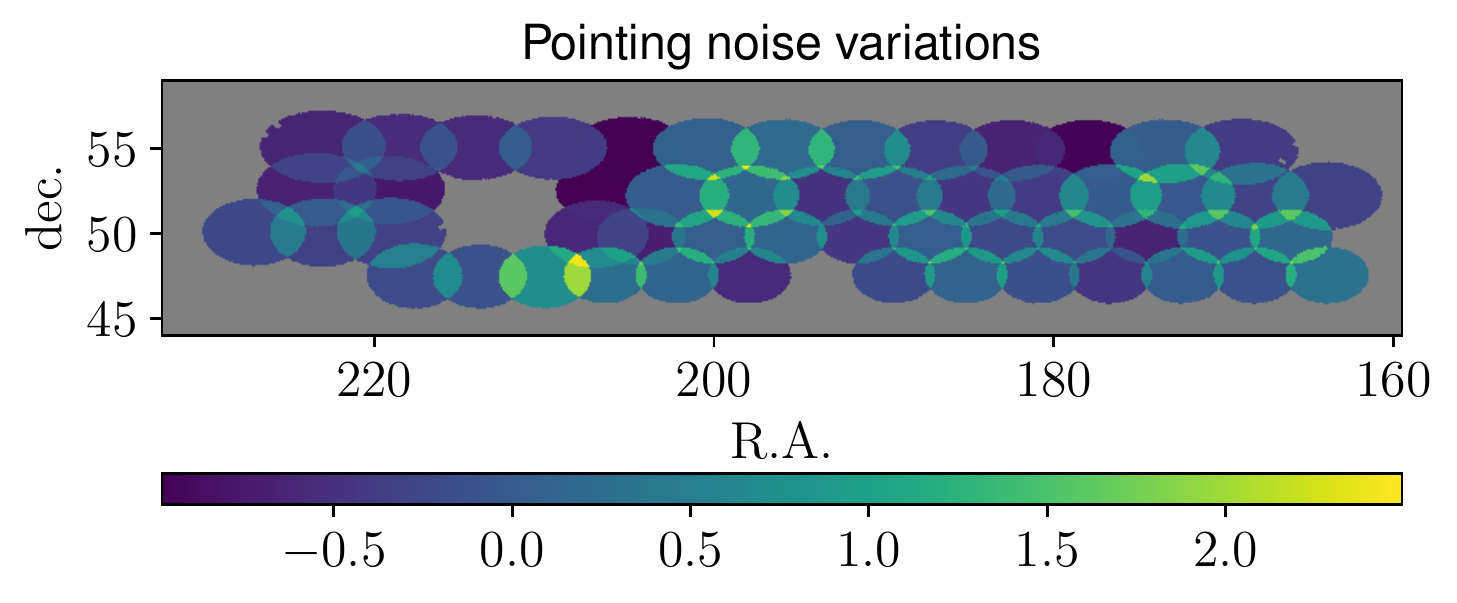}
        \includegraphics[width=0.47\textwidth]{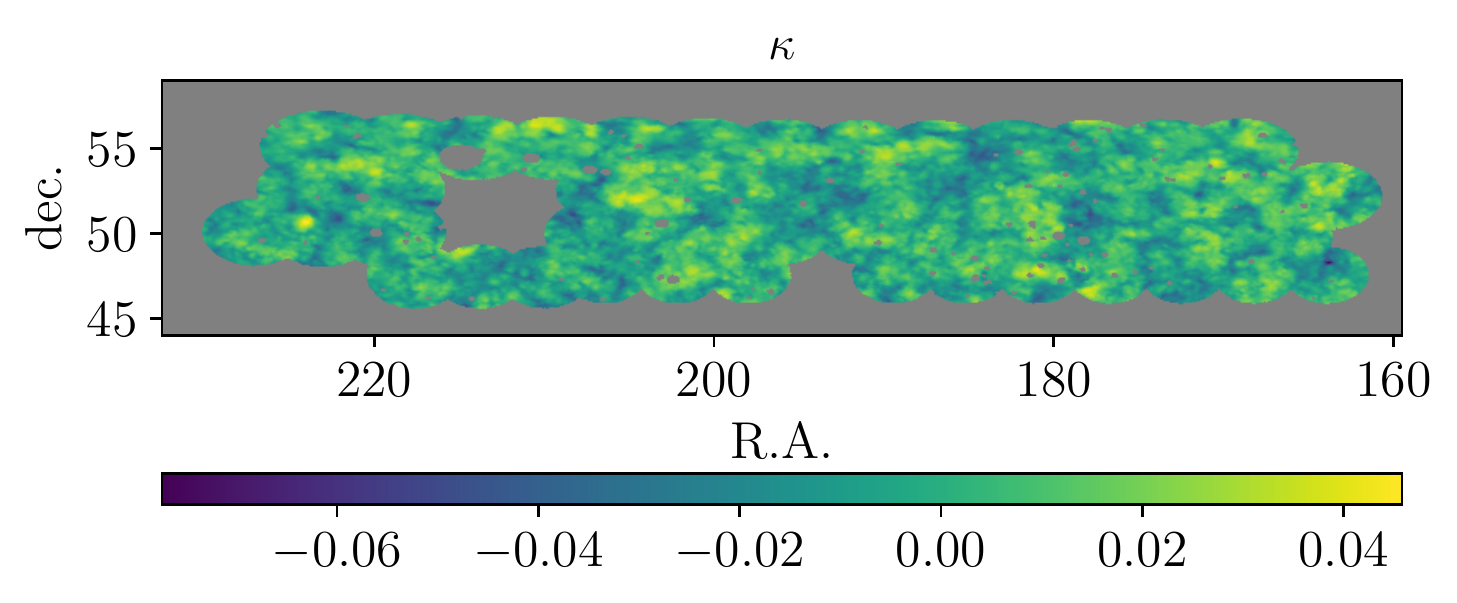}
        \includegraphics[width=0.47\textwidth]{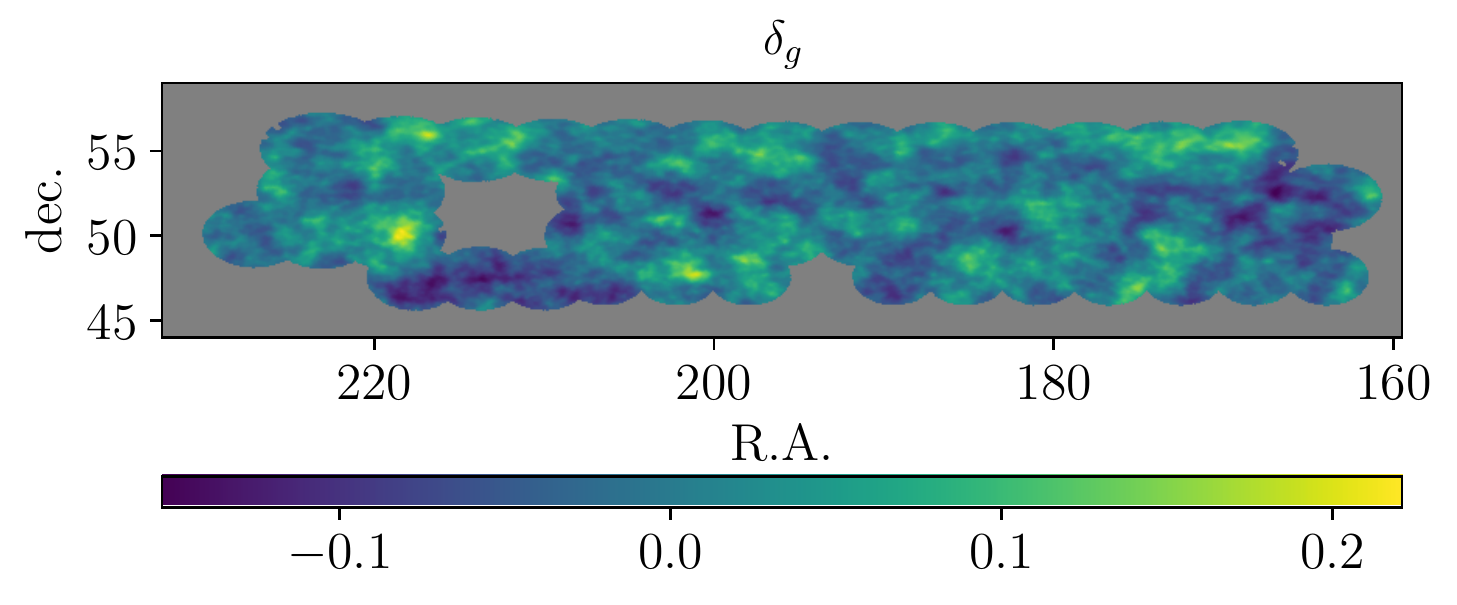}
        \caption{Various maps used in the analysis. {\sl Top left:} binary mask for the LoTSS DR1 sample. {\sl Top right:} $S_{\rm cut}=2\,\mJy$ mean density map. {\sl Middle left:} relative fluctuations in the noise rms. {\sl Middle right:} Wiener-filtered CMB lensing convergence map. {\sl Bottom:} galaxy overdensity map smoothed with the same filter.}\label{fig:maps}
      \end{figure*}
      A crucial aspect of any analysis involving the fluctuations in galaxy number counts $\delta_g$, is estimating the expected mean number of objects in each sky pixel. This involves accurately characterizing the geometry of the observed footprint, as well as the expected fluctuations in the number of observed sources due to variations in image noise. We will label these two the \emph{mask} and \emph{mean density} map respectively.
      
      \subsubsection{Sky mask}\label{sssec:methods.weights.masks}
        We produce masks for each of the 58 pointings from their associated low-resolution images at resolution $N_{\rm side}=2048$. Each mask is a binary sky map containing 0s or 1s depending on whether the center of a given pixel lies inside the pointing. The final mask is then the logical addition of the 58 pointing masks.
        
        When computing power spectra with the lower resolution maps, the associated mask is computed by averaging down the $N_{\rm side}=2048$ masks. The resulting low-resolution mask contains, in each pixel, the fraction of its area that was observed. This mask is shown in the top left panel of Fig. \ref{fig:maps}.
        
      \subsubsection{Mean density}\label{sssec:methods.weights.meandens}
        The expected mean number of observed objects in a given pixel varies across the sky as a result of spatial variations in observing conditions, resulting on fluctuations in flux noise. If not accounted for (i.e. if interpreted as intrinsic fluctuations in the galaxy number density), they will bias the estimated two-point functions and, through them, the final parameter constraints. We used a semi-analytic method to reconstruct the expected fluctuations in the mean density for a given sample that uses the same noise model used by \citet{2019arXiv190810309S}.
        
        Let $S_{\rm obs}$ be the observed flux of a given object with true flux $S_{\rm true}$. The probability to detect objects with observed fluxes above a given threshold $S_{\rm thr}$ is:
        \begin{align}\nonumber
          P(>S_{\rm thr})
          &=\int_{0}^\infty dS_{\rm true}\,p(S_{\rm true})\,\Pi(S_{\rm true}, S_{\rm thr})
        \end{align}
        where $p(S_{\rm true})$ is the distribution of true fluxes, and $\Pi(S_{\rm true},S_{\rm thr})$ is the probability for a source with true flux $S_{\rm true}$ to have an observed flux above $S_{\rm thr}$, given in terms of the conditional distribution of observed fluxes $p(S_{\rm obs}|S_{\rm true})$ as
        \begin{equation}
          \Pi(S_{\rm true},S_{\rm thr})\equiv\int_{S_{\rm thr}}^\infty dS_{\rm obs}\,p(S_{\rm obs}|S_{\rm true}).
        \end{equation}
        Assuming the noise in the measured fluxes has a Gaussian distribution with standard deviation $\sigma_S$ \citep[as done in][]{2019arXiv190810309S}, this is simply
        \begin{equation}
          \Pi(S_{\rm true},S_{\rm thr})=\frac{1}{2}{\rm erfc}\left[\frac{S_{\rm true}-S_{\rm thr}}{\sqrt{2}\sigma_N}\right],
        \end{equation}
        where ${\rm erfc}$ is the complementary error function.
      
        We therefore need two ingredients in order to compute $P(>S_{\rm thr})$:
        \begin{itemize}
          \item An estimate of the noise rms in each pixel $\sigma_N(\nv)$. As done in \citet{2019arXiv190810309S}, we make histograms of the rms noise for all sources in a given pixel, and use them to produce maps of the mean and median rms noise per pixel. The rms noise of each source is given as the averaged background rms of the corresponding island. Note that, in order to have enough galaxies in each pixel on average, we produce these maps with resolution $N_{\rm side}=256$.
          \item An estimate of the distributon of true fluxes $p(S_{\rm true})$. We obtain this from the SKADS semi-empirical simulation \citep{2008MNRAS.388.1335W}. We queried the SKADS database to retireve the flux distribution at $151\,{\rm MHz}$, which we scaled to $144\,{\rm MHz}$ assuming a spectral index $\alpha=-0.7$.
        \end{itemize}
        
        We use this method to produce a mean density map for our sample. Since we only consider sources detected above $5\sigma$, the threshold flux is $S_{\rm thr}={\rm max}(5\sigma_N,S_{\rm cut})$, where $S_{\rm cut}$ is the flux cut defining the sample. Figure \ref{fig:maps} shows, in the top right panel, the resulting map for our fiducial $S_{\rm cut}=2\mJy$. The sample is highly homogeneous, with a mean density $\bar{n}=0.12\,{\rm arcmin}^{-2}$ that varies by $\sim1.4\%$ across the footprint (1 standard deviation).

      \subsubsection{Noise variations}\label{sssec:methods.weights.surprop}
        The number of detected sources is likely to depend strongly on the depth of a given sky region. As a tracer for those depth variations, we map the coadded inverse background noise variance. For each pointing, we estimate its noise rms $\sigma_N$ as the standard deviation of the low-resolution residual map made available with the DR1. The final inverse noise variance map is generated by adding the values of $\sigma_N^{-2}$ for all pointings overlapping on each pixel. It is worth noting that the procedure used to generate this map does not reflect exactly the mosaicing process used in \lotss{}. Additionally, there are probably fluctuations of the flux density scale between pointings that have not been quantified. It is expected that these shortcomings will improve with future data releases. This map is shown in middle left panel of Figure \ref{fig:maps}, and will be used in the next section to study the impact of this systematic on the power spectrum.

    \subsection{Maps and power spectra}\label{ssec:methods.cls}
      Before computing the power spectra, we first generate a map of the overdensity field $\delta_g(\nv)$ as:
      \begin{equation}
        \delta_g(\nv)=\frac{N(\nv)}{\bar{N}w_g(\nv)}-1,
      \end{equation}
      where $N(\nv)$ is the number of galaxies observed in the pixel with position $\nv$, and $w_g(\nv)$ is a weight map given by the product of the sky mask described in Section \ref{sssec:methods.weights.masks}, and the mean density map described in Section \ref{sssec:methods.weights.meandens}. $w_g(\nv)$ therefore quantifies the fraction of galaxies passing our cuts at a given pixel that should have been observed. The quantity $\bar{N}$ is the mean number of objects per pixel in the sample, estimated as $\bar{N}=\langle N(\nv)\rangle_\theta/\langle w_g(\nv)\rangle_\theta$, where $\langle\cdot\rangle_\theta$ denotes an average over all pixels in the map. Finally, to avoid using heavily masked pixels, we set to zero all pixels in $\delta_g(\nv)$ and $w_g(\nv)$ where $w_g(\nv)<0.5$.

      Once we have the two maps $\delta_g(\nv)$ and $\kappa(\nv)$ (both shown in Fig. \ref{fig:maps}, and their associated masks $w_g(\nv)$ and $w_\kappa(\nv)$ (the latter described in Section \ref{ssec:data.planck}), we compute the auto-correlation of $\delta_g$ ($C_\ell^{gg}$), and its cross-correlation with $\kappa$ ($C_\ell^{g\kappa}$) using the so-called \emph{pseudo-$C_\ell$} estimator \citep{1973ApJ...185..413P,2002ApJ...567....2H} using the {\tt NaMaster} code\footnote{\url{https://github.com/LSSTDESC/NaMaster}} \citep{2019MNRAS.484.4127A}. The pseudo-$C_\ell$ algorithm uses analytical methods to calculate the coupling of different multipoles $\ell$ due to the incomplete sky coverage. Although the method is in principle less optimal than minimum variance quadratic estimators, it is able to achieve almost equivalent uncertainties for sufficiently flat power spectra (as is the case here) with a much higher computational speed. We bin both power spectra in bands of $\Delta\ell=50$ starting at $\ell=2$.
      
      The estimated overdensity map contains shot noise due to the discrete nature of galaxy number counts. The associated contribution to the angular power spectrum (commonly called the ``noise bias''), can be estimated analytically as described in \citet{2019MNRAS.484.4127A}. We will assess the validity of this calculation in Section \ref{ssec:results.cls}.
      
      We also use {\tt NaMaster} to estimate the covariance matrix of the power spectra following \citet{2019JCAP...11..043G}. This is calculated as the so-called \emph{Gaussian covariance}, which approximates both $\kappa$ and $\delta_g$ as Gaussian random fields. The covariance matrix estimator implemented in {\tt NaMaster} accurately accounts for the correlation between different $\ell$ bins induced by the incomplete sky coverage. Since the estimator requires a best-guess estimate of the underlying power spectra ($C_\ell^{gg}$, $C_\ell^{g\kappa}$ and $C_\ell^{\kappa\kappa}$), the covariance is estimated in two steps: first, we compute theoretical power spectra for cosmological parameters fixed to the best-fit values found by \planck{} \citep{2018arXiv180706209P} and a constant galaxy bias $b_g=1.3$ assuming the LoTSS redshift distribution, which provides a good visual fit to the data. The resulting covariance is then used in the likelihood described in Section \ref{ssec:methods.like} to find the best-fit parameters. These are used to estimate new theory power spectra that are then used by {\tt NaMaster} to estimate our final covariance matrix. Note that the auto-spectra $C^{gg}_\ell$ and $C^{\kappa\kappa}_\ell$ should contain both the signal and noise contributions. For $C^{gg}_\ell$ we use the shot noise component described above, while for $C^{\kappa\kappa}_\ell$ we use the noise curves provided in the \planck{} data release.
      
      It is worth noting that, although the Gaussian covariance described above assumes Gaussian statistics for both $\delta_g$ and $\kappa$, which is known to be inaccurate due to the non-linear growth of structure, the covariance estimated using this method accounts for the largest fraction of the statistical uncertainty \citep{2018JCAP...10..053B,2020JCAP...03..044N}, and is an excellent approximation in the range of scales studied here.

    \subsection{Likelihood}\label{ssec:methods.like}
      In order to extract information from the measured $C^{gg}_\ell$ and $C^{g\kappa}_\ell$ we use a Gaussian likelihood of the form:
      \begin{equation}
        \chi^2\equiv-2\log p({\bf d}|{\bf q})=({\bf d}-{\bf t}({\bf q}))^T{\sf Cov}^{-1}({\bf d}-{\bf t}({\bf q})),
      \end{equation}
      where the data vector ${\bf d}$ denotes all measured power spectra and ${\bf t}({\bf q})$ is the theoretical prediction for ${\bf d}$ given a set of parameters ${\bf q}$.
      
      We will present results for different choices of data vector and parameters. Specifically, we will present constraints based on $C_\ell^{gg}$ and $C_\ell^{g\kappa}$ alone, as well as from the combination of both. We will also explore different combinations of three free parameters:
      \begin{itemize}
        \item The galaxy bias $b_g$ (within the two redshift evolution models described in Section \ref{ssec:theory.cls}).
        \item $z_{\rm tail}$, which quantifies the extent of the redshift distribution tail when parametrized according to Eq. \ref{eq:nz_ana}.
        \item The amplitude of matter fluctuations as parametrized by $\sigma_8$.
      \end{itemize}
      We fix all cosmological parameters (including $\sigma_8$ when not used as a free parameter) to the best-fit values found by \planck{} \citep{2018arXiv180706209P}.
      
      We will only use the multipoles $\ell$ smaller than $\ell_{\rm max}=500$, corresponding to a wavenumber $k_{\rm max}\simeq0.15\,{\rm Mpc}^{-1}$ at $z\simeq1$. Thus we make sure that we only use modes where a linear, scale-dependent bias relation is a good approximation, and where the non-Gaussian contributions to the covariance matrix can be neglected.

  \section{Results}\label{sec:results}
    \subsection{Power spectra and covariances}\label{ssec:results.cls}
      \begin{figure}
        \centering
        \includegraphics[width=0.49\textwidth]{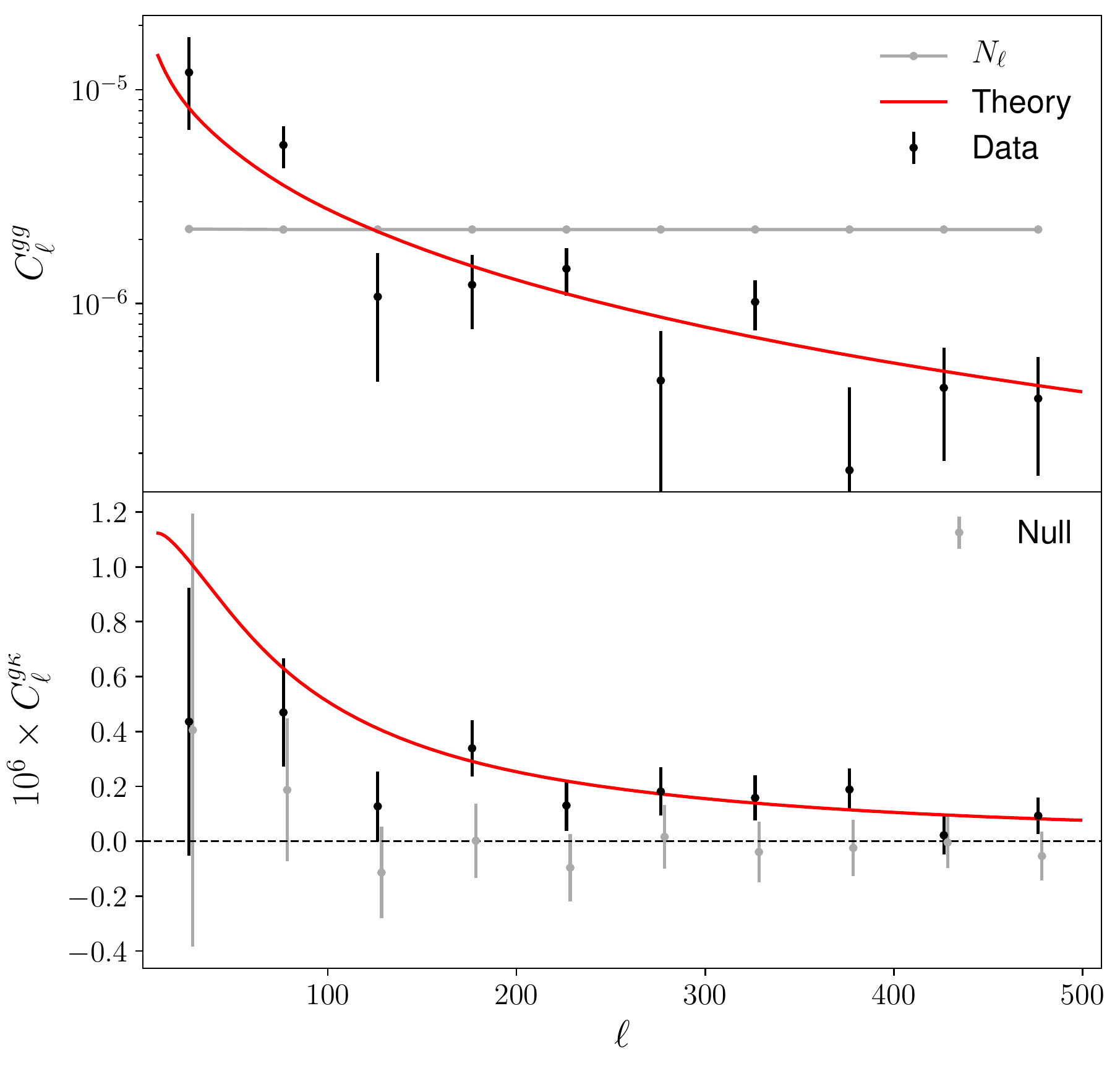}
        \caption{Galaxy auto-correlation $C^{gg}_\ell$ (top panel) and its cross-correlation with the CMB lensing convergence $C_\ell^{g\kappa}$ (bottom panel). The measured power spectra are shown as black points with error bars. The solid red curve shows the best-fit theory prediction assuming the SKADS redshift distribution, the \planck{} cosmological parameters, and a galaxy bias that grows inversely with the linear growth factor (Eq. \ref{eq:b_grth}), with an amplitude $b_g=2.1$. The noise bias due to shot noise in the auto-correlation is shown as a gray line in the top panel. The gray points and error bars in the bottom panel show a null cross-correlation calculated by correlating the galaxy overdensity map with a randomly rotated CMB convergence map.}\label{fig:cls_all}
      \end{figure}
      \begin{figure}
          \centering
          \includegraphics[width=0.49\textwidth]{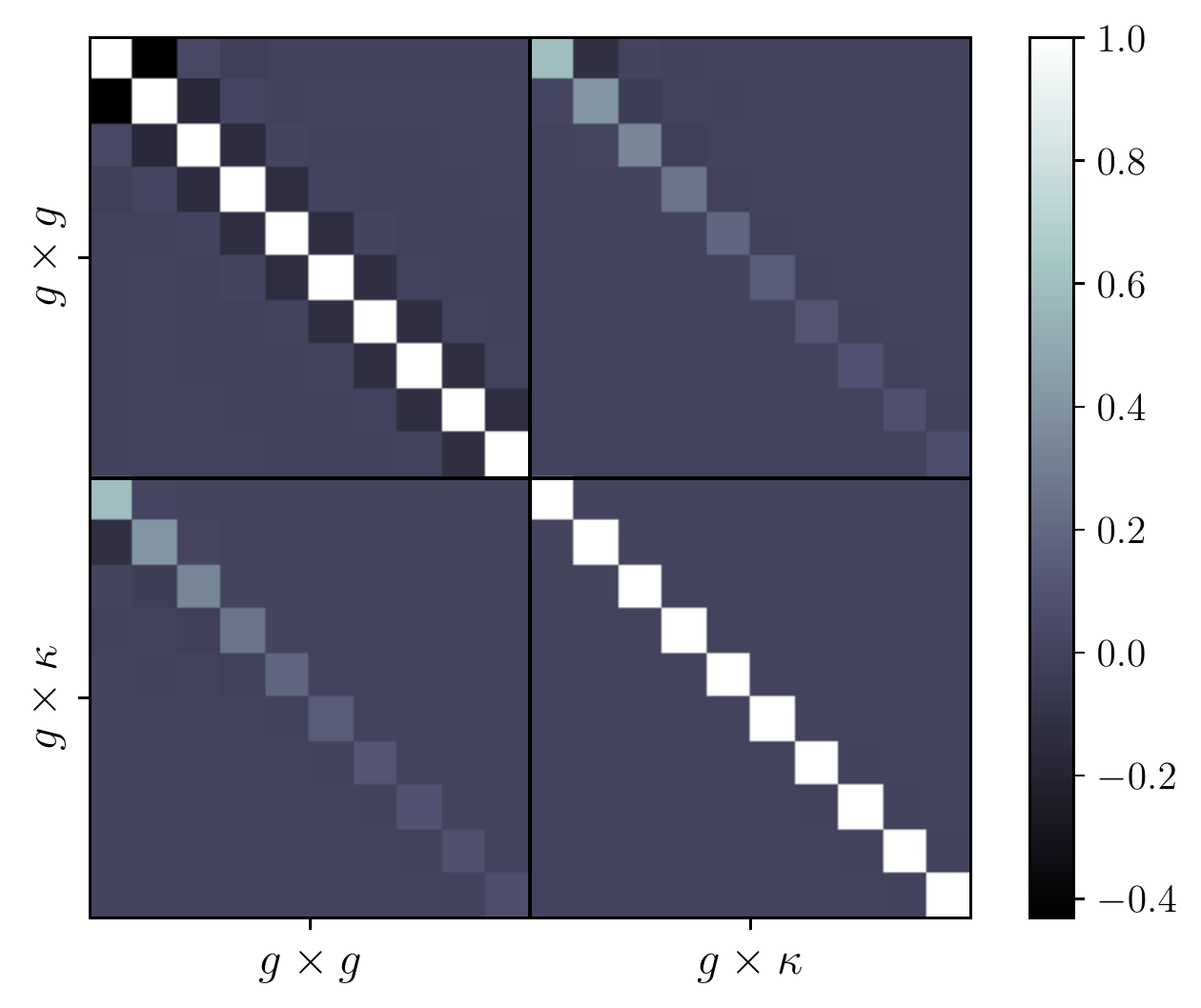}
          \caption{Correlation matrix associated with the covariance matrix used in this analysis. The covariance was calculated analytically using the methods in \citet{2019JCAP...11..043G}. The auto- and cross-correlation are $\sim50\%$ correlated with each other on the largest scales.}
          \label{fig:cov}
      \end{figure}

      The power spectra $C^{gg}_\ell$ and $C^{g\kappa}_\ell$, measured using the methods described in Section \ref{ssec:methods.cls}, are shown as black points with error bars in the upper and lower panels of Figure \ref{fig:cls_all}. The correlation matrix ${\sf r}_{ij}={\sf Cov}_{ij}/\sqrt{{\sf Cov}_{ii}{\sf Cov}_{jj}}$ associated with the covariance matrix of the full data vector is shown in Fig. \ref{fig:cov}. The lowest multipoles of $C_\ell^{gg}$ and $C_\ell^{g\kappa}$ are $\sim40\%$ correlated. 

\begin{figure}
        \begin{center}
          \includegraphics[width=0.47\textwidth]{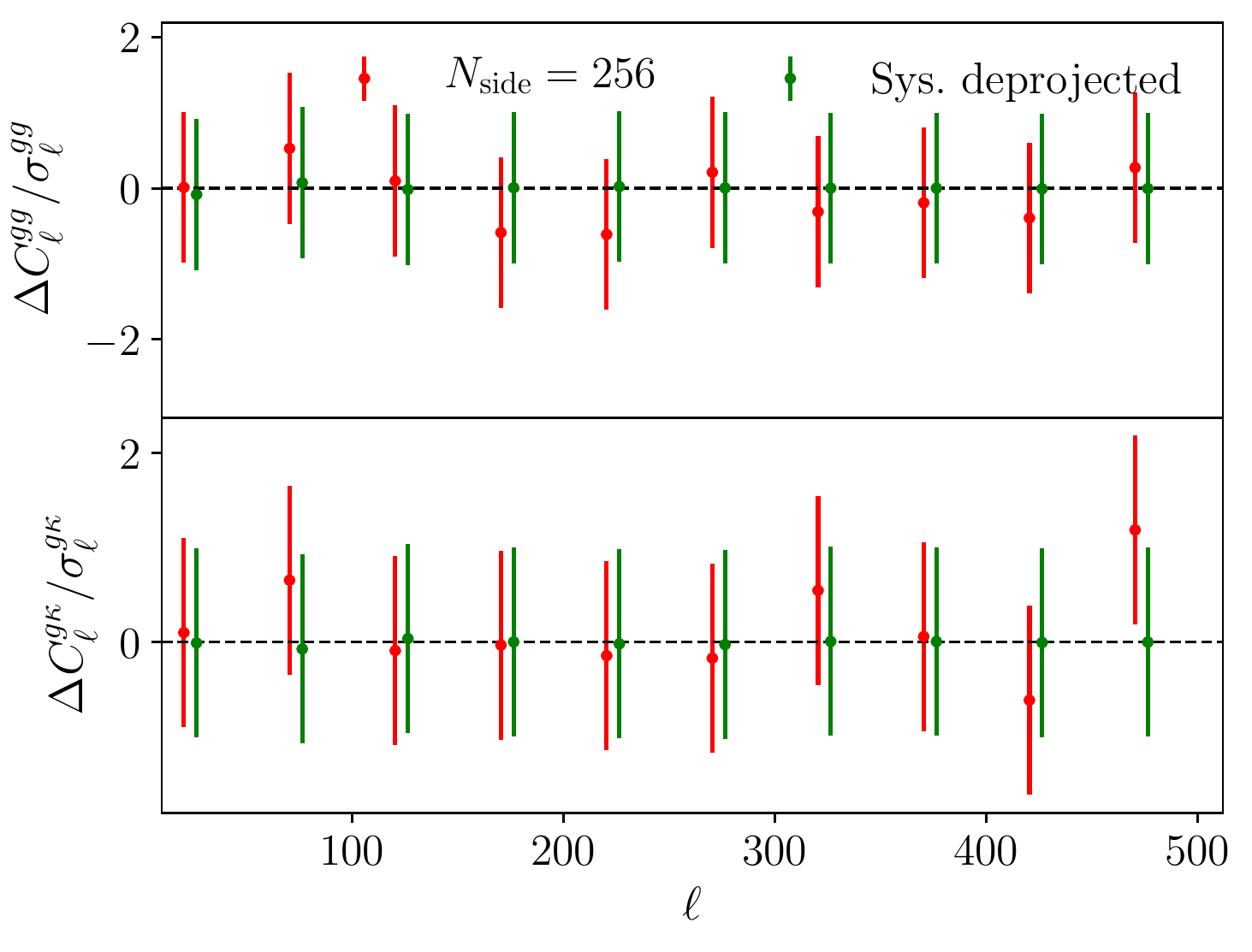}
          \caption{Difference between the fiducial auto- and cross-spectra (top and bottom panels respectively) and two alternative measurements as a fraction of the 1$\sigma$ uncertainties. The red points show the effect of using coarser pixels ($N_{\rm side}=256$). The green points show the result of deprojecting the fluctuations in the mean density map and the noise variations map (described in Sections \ref{sssec:methods.weights.meandens} and \ref{sssec:methods.weights.surprop}) from the galaxy overdensity map, as proxies for potential contaminants of the observed $\delta_g$.}\label{fig:syst_cl}
        \end{center}
      \end{figure}

      Before using these measurements to extract parameter constraints, we perform a number of sanity checks and null tests. The main concern regarding $C^{gg}_\ell$ is the presence of residual systematics in the galaxy overdensity producing extra power on large scales \citep{2012MNRAS.424..564R,2013MNRAS.435.1857L,2016ApJS..226...24L}. In our case, the most likely source of such systematics is fluctuations in survey completeness caused by variations in survey properties. As a proxy for these residual systematics, we use the maps of pointing noise variation and the fluctuations in the mean density map itself. We then compare the power spectrum $\hat{C}_\ell^{gg}$ estimated from the original overdensity map and from a ``systematics-deprojected'' map, in which these systematic templates are projected out at the map level \citep[see][for details]{2019MNRAS.484.4127A}.  We also verify that the power spectra are stable against the choice of pixelization by recomputing them with a resolution parameter $N_{\rm side}=256$. The result of these two tests is displayed in Fig. \ref{fig:syst_cl}. The figure shows the relative difference between the fiducial and alternative power spectra as a fraction of the 1$\sigma$ uncertainties. The power spectra are robust to the choice of pixelization and to the potential contamination from variations in survey completeness within the range of scales explored here.
      
      In order to validate the cross-correlation $\hat{C}_\ell^{g\kappa}$, and to test for a potential mis-estimate of the statistical uncertainties, we perform one further null test by cross correlating the overdensity map $\delta_g$ with a random convergence map generated by arbitrarily rotating the original \planck{} map. The measured cross correlation is shown by the grey points in the lower panel of Fig. \ref{fig:cls_all}. Fitting a constant amplitude to these measurements we find the best-fit value $A_{\rm null}=(-2.9\pm4.1)\times10^{-8}$, consistent with zero.
      
      The galaxy auto-correlation $\hat{C}_\ell^{gg}$ could still be affected by unknown systematics that are not well described by the two template maps used above. These systematics would typically affect the largest scales, comparable with the size of a pointing ($\ell\sim\pi/\delta\theta\sim50$), and therefore we can test for the relevance of unknown contaminants by removing those scales from the analysis. Another source of systematic uncertainty that could bias our results is a mis-estimate of the noise bias in $\hat{C}_\ell^{gg}$. This noise bias is shown as a gray solid line in the top panel of Fig. \ref{fig:cls_all}, and was calculated analytically. An error in this estimate would mostly affect the smallest scales, which are dominated by shot noise, potentially biasing the inferred parameters. To study the impact of these two systematics, we constrain the galaxy bias parameter $b_g$ from the auto-correlation $\hat{C}_\ell^{gg}$ alone in three cases:
      \begin{enumerate}
          \item A fiducial case, using all the data and the fiducial Gaussian covariance matrix. We find a bias value $b_g=2.10\pm0.10$.
          \item We remove the first bandpower, covering scales $\ell<50$. The inferred bias is $b_g=1.98\pm0.12$.
          \item Instead of analytically computing the noise bias and subtracting it from the data, we marginalize over a constant offset to $\hat{C}_\ell^{gg}$ with a free amplitude parameter. This is done analytically by simply modifying the inverse covariance as \citep{1992ApJ...398..169R}:
          \begin{equation}
            {\sf Cov}^{-1}\rightarrow{\sf Cov}^{-1}-\frac{({\sf Cov}^{-1}{\bf n})({\sf Cov}^{-1}{\bf n})^T}{{\bf n}^T{\sf Cov}^{-1}{\bf n}},
          \end{equation}
          where ${\bf n}$ is our fiducial estimate of the noise bias. The corresponding constraint is $b_g=2.23\pm0.16$.
      \end{enumerate}
      In all cases, we parametrize the redshift dependence of $b_g(z)$ using the ``constant amplitude'' model, and we assume the SKADS redshift distribution. Given the restricted range of scales used in this analysis, marginalizing over the noise bias leads to a $60\%$ increase in the final uncertainties. Nevertheless, the inferred values of $b_g$ are compatible with each other in all cases, and therefore we conclude that the measured spectra are robust to unknown sky contaminants and uncertainties in the shot noise level.

    \subsection{Cross-correlation significance}\label{ssec:results.detect}
      \begin{figure*}
        \begin{center}
          \includegraphics[width=0.49\textwidth]{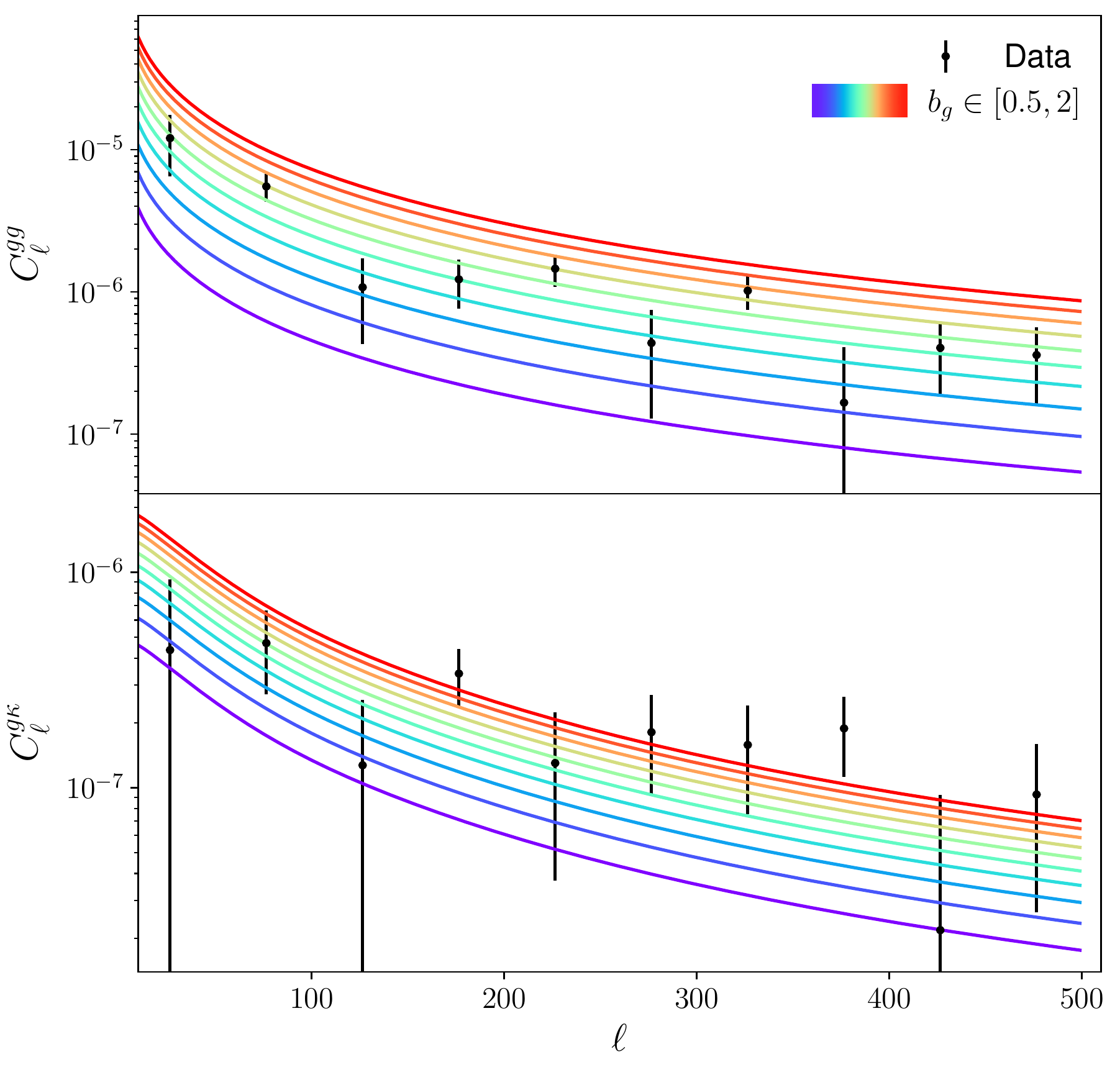}
          \includegraphics[width=0.49\textwidth]{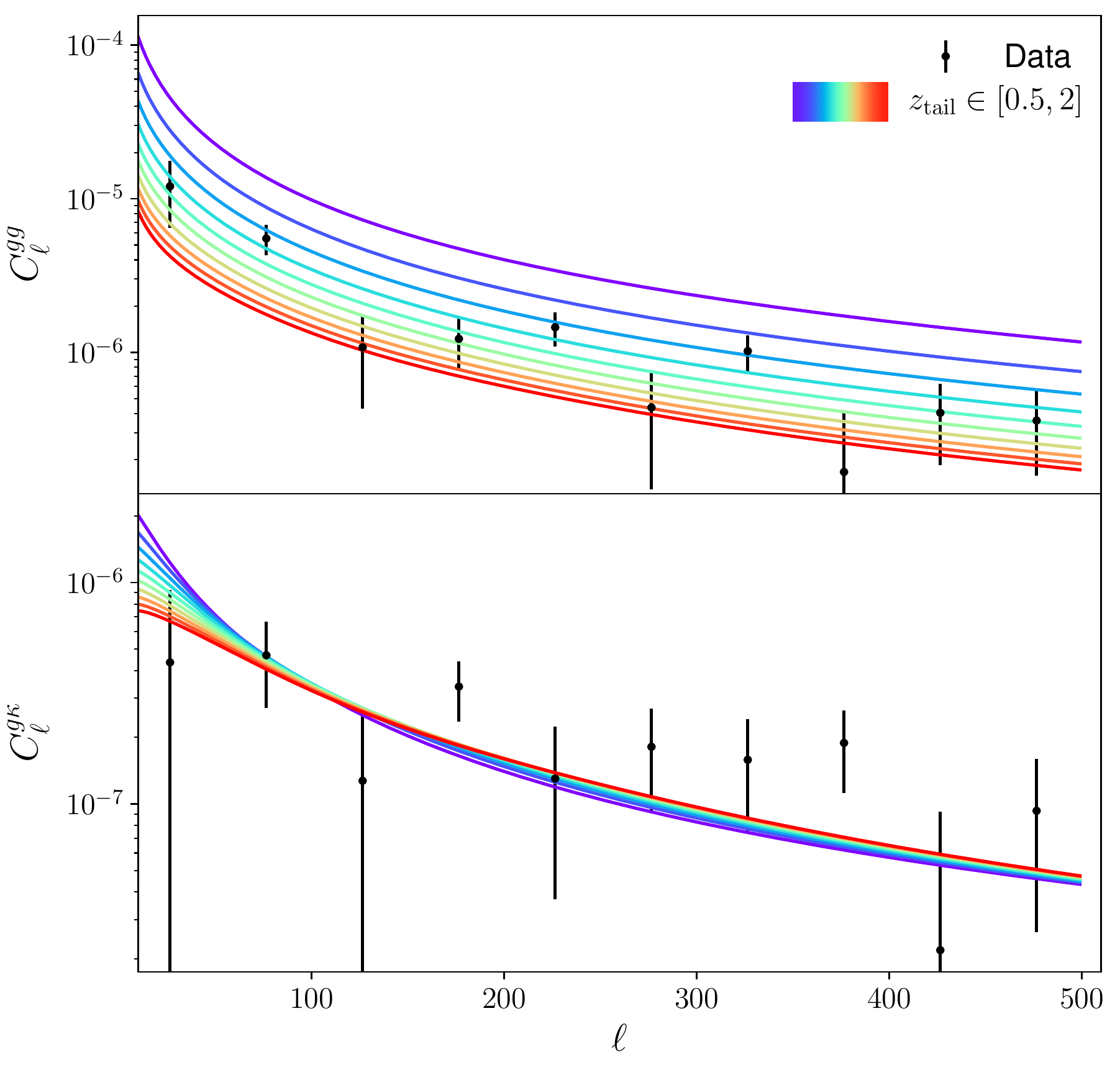}
          \caption{Measured auto- and cross-correlation (black dots with error bars int the top and bottom panels respectively), together with the theory prediction for different values of the galaxy bias $b_g$ and the high redshift tail $z_{\rm tail}$ (left and right panels respectively), both in the range $(0.5, 2.0)$. A ``constant amplitude'' model is assumed for the redshift evolution of the galaxy bias. $b_g$ is fixed to 1.3 in the right panel, while $z_{\rm tail}=1.1$ in the left one. While both $b_g$ and $z_{\rm tail}$ affect the amplitude of the auto-correlation, the cross-correlation depends only mildly on the high-redshift tail, making it possible to break the degeneracy between both parameters by combining $C_\ell^{gg}$ and $C_\ell^{g\kappa}$.}\label{fig:cl_intuit}
        \end{center}
      \end{figure*}
      The measured cross-correlation $\hat{C}_\ell^{g\kappa}$ is shown in the lower panel of Figure \ref{fig:cls_all}. We can quantify the significance of this detection in two different ways. First, the probability to exceed associated with the $\chi^2$ of the measurement of $\hat{C}^{g\kappa}_\ell$ with respect to the null hypothesis (${\bf t}=0$) is $p(>\chi^2)=8\times10^{-5}$, corresponding to a $\sim 4\sigma$ detection.
      
      Secondly, we fit the measured cross-correlation to a single parameter model of the form
      \begin{equation}
        \hat{C}_\ell^{g\kappa}=b_g C_\ell^{g\kappa}|_{b=1},
      \end{equation}
      where $C_\ell^{g\kappa}|_{b=1}$ is the predicted signal assuming the \planck{} best-fit cosmological parameters, the SKADS redshift distribution and a unit galaxy bias parameter $b_g=1$ in the constant-amplitude redshift dependent model. The constraint on the free amplitude $b_g$ is $b_g=1.59\pm 0.31$, corresponding to a $5.2\sigma$ measurement of the cross-correlation. Note that this is mathematically equivalent to quantifying the significance as the square root of the difference in $\chi^2$ between the null hypothesis and this best-fit model $\sqrt{\Delta\chi^2}$.

    \subsection{Galaxy bias and high-redshift tails}\label{ssec:results.bias_tail}
      \begin{figure}
        \begin{center}
          \includegraphics[width=0.49\textwidth]{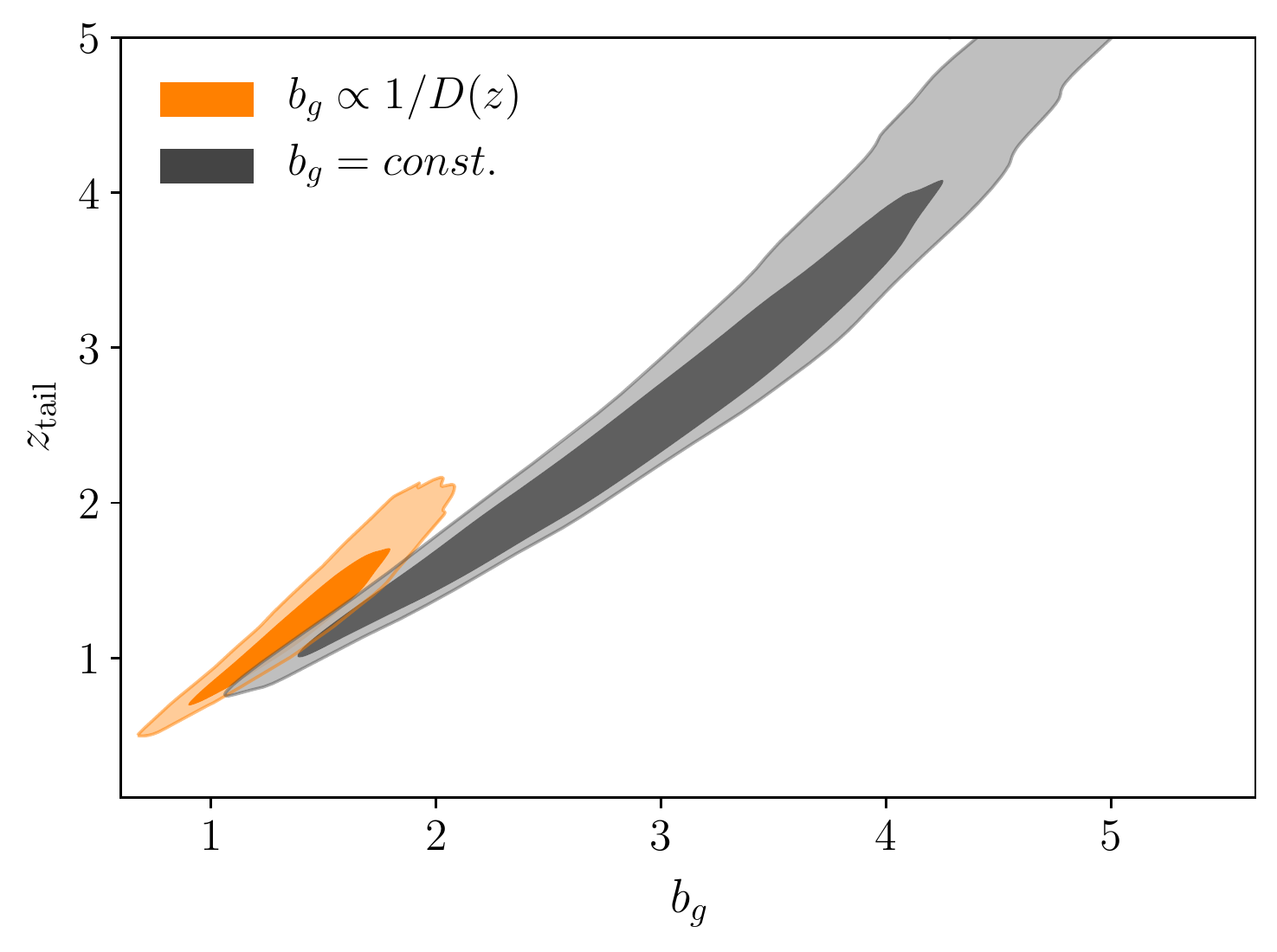}
          \caption{Constraints on galaxy bias $b_g$ and the high-redshift tail parameter $z_{\rm tail}$ from the combination of $C_\ell^{gg}$ and $C_\ell^{g\kappa}$. Results are shown for a constant galaxy bias as a function of redshift, the ``constant bias'' model (black), and for a bias that scales with the inverse of the linear growth factor, the ``constant amplitude'' model (orange).}\label{fig:ztail_nz}
        \end{center}
      \end{figure}
      As described in Section \ref{ssec:data.nz}, we have at our disposal at least three different estimates of the redshift distribution for our sample, which differ mostly in terms of the size of their high-redshift tails. It is therefore important to understand the level to which the galaxy bias constraints depend on the redshift distribution uncertainties. Assuming the \planck{} cosmological parameters and a constant-amplitude redshift dependent model for the galaxy bias, the constraints on $b_g$ assuming these three different redshift distributions, and using both $\hat{C}_\ell^{gg}$ and $\hat{C}_\ell^{g\kappa}$ are:
      \begin{align}\nonumber
        &b_g=1.12\pm 0.05,\hspace{6pt}\textrm{(LoTSS-VAC)},\\
        &b_g=1.84\pm 0.09,\hspace{6pt}{\rm (VLA-COSMOS)},\\\nonumber
        &b_g=2.10\pm 0.10,\hspace{6pt}{\rm (SKADS)}.
      \end{align}
      While the VLA-COSMOS and SKADS estimates are roughly compatible with each other, they differ from the LoTSS-VAC result by  more than $7\sigma$. The differences between the different redshift distribution estimates are therefore highly significant in this context. In order to investigate this further, we repeat the exercise using only the CMB lensing cross-correlation. In this case the constraints are fully compatible with each other (albeit with larger error bars):
      \begin{align}\nonumber
        &b_g=1.35\pm 0.25,\hspace{6pt}{\rm (LoTSS)},\\
        &b_g=1.46\pm 0.28,\hspace{6pt}{\rm (VLA-COSMOS)},\\\nonumber
        &b_g=1.59\pm 0.31,\hspace{6pt}{\rm (SKADS)}.
      \end{align}
      
      The reason for the disagreement between the different inferred bias values with the full data vector is the well-known fact that both the galaxy bias and the width of the redshift distribution have a degenerate effect on the amplitude of the galaxy auto-correlation. Structure is washed out in samples with broader redshift distributions, which can be compensated with a larger $b_g$.
      \begin{figure}
        \begin{center}
          \includegraphics[width=0.49\textwidth]{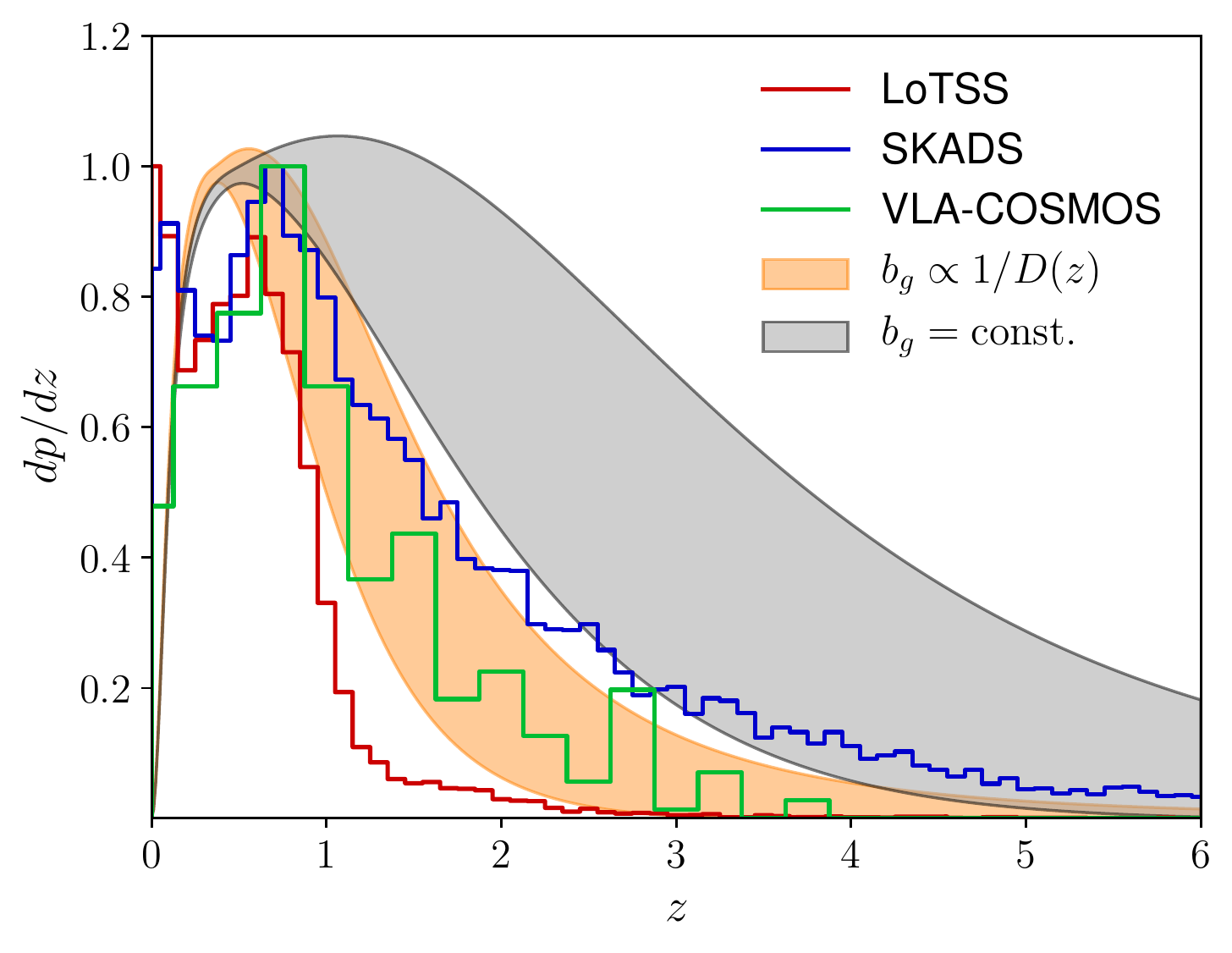}
          \caption{1$\sigma$ constraints on the redshift distribution of the LoTSS flux-limited sample obtained from the combination of $C_\ell^{gg}$ and $C_\ell^{g\kappa}$. Results are shown for a constant bias model (gray region), and for  a bias that grows with the inverse of the linear growth factor (orange region). The redshift distributions inferred from the photometric redshifts in the LoTSS value-added catalog, from the 3 GHz VLA-COSMOS catalog, and from the SKADS simulations are shown in red, green and blue respecitvely for comparison. All redshift distribution are normalized to have the same maximum amplitude. The constraints show a preference for longer redshift tails than the one predicted by the LoTSS photo-$z$s.}\label{fig:nz_tail}
        \end{center}
      \end{figure}
      \begin{figure}
        \begin{center}
          \includegraphics[width=0.49\textwidth]{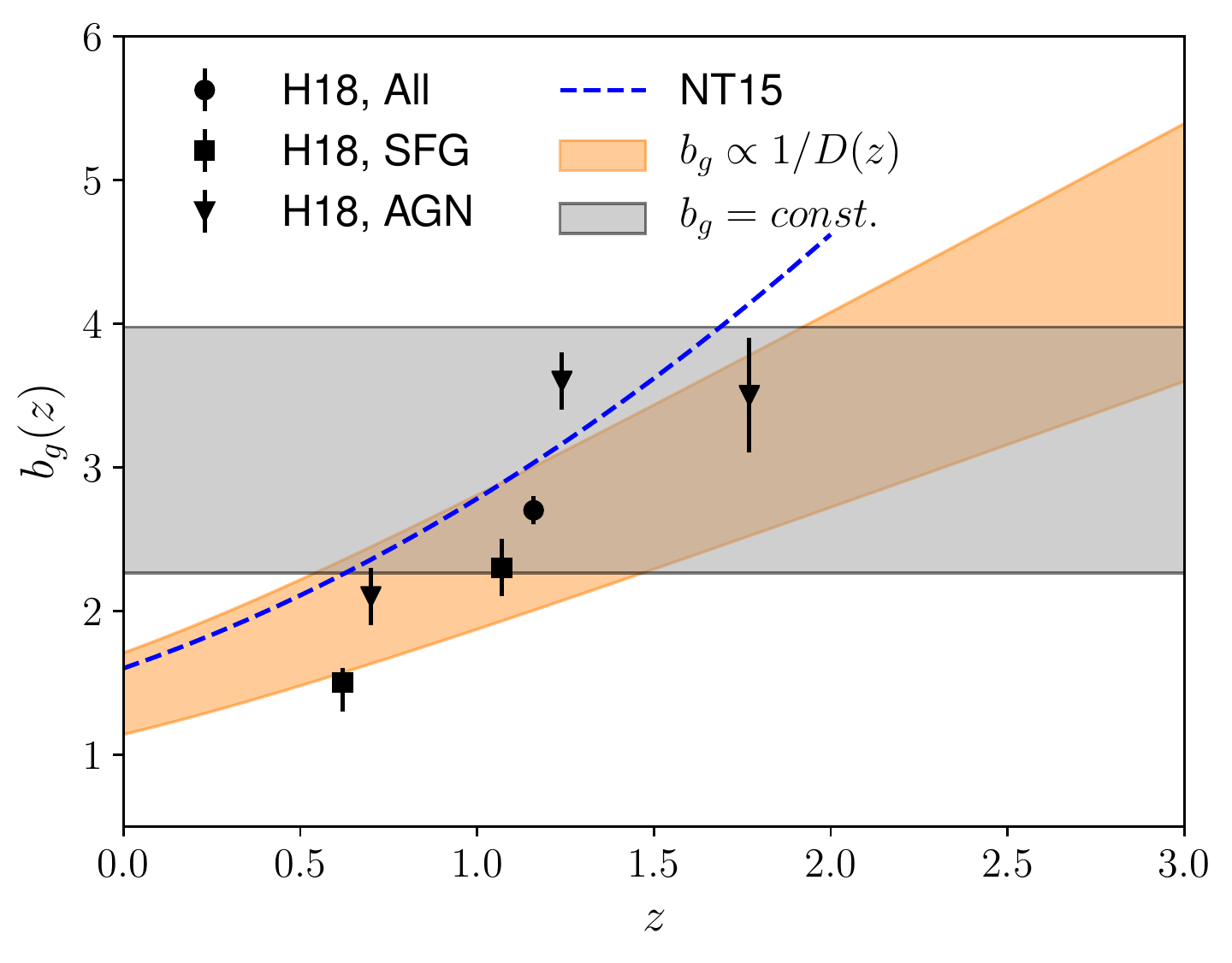}
          \caption{1$\sigma$ constraints on the galaxy bias $b_g(z)$ from the combination of $C_\ell^{gg}$ and $C_\ell^{g\kappa}$. Results are shown for a constant bias model (gray region), and for  a bias that grows with the inverse of the linear growth factor (orange region). The figure also shows measurements of the bias for different radio galaxy samples in \citet{Hale:2017wub} (black symbols with error bars), and for the NVSS sample of \citet{2015ApJ...812...85N}.}\label{fig:bzs}
        \end{center}
      \end{figure}
      
      To understand the role played by both parameters in the auto- and cross-correlation, Figure \ref{fig:cl_intuit} shows the measured power spectra together with the theoretical predictions for varying values of the galaxy bias $b_g$ (left panel) and the high-redshift tail $z_{\rm tail}$ (right panel). While the galaxy bias affects the amplitude of the auto- and cross-correlations (quadratically and linearly respectively), an increasing high-redshift tail lowers the auto-correlation while leaving the cross-correlation almost constant. We can understand the latter result as follows: since the CMB lensing kernel $W_\kappa$ extends to very high redshifts, a variation in the width of the galaxy redshift distribution leaves the overlap between $W_g$ and $W_\kappa$ almost unchanged. On the one hand, this makes the cross-correlation between CMB lensing and galaxy clustering robust to uncertainties in the redshift distribution width. On the other hand, the different response of the auto- and cross-correlation to $z_{\rm tail}$ should allow us to break its degeneracy with $b_g$, allowing us to simultaneously constrain both parameters by combining $\hat{C}_\ell^{gg}$ and $\hat{C}_\ell^{g\kappa}$.
      
      To explore this, we parametrize the redshift distribution according to Eq. \ref{eq:nz_ana} and derive constraints on two free parameters, $b_g$ and $z_{\rm tail}$, with flat priors $b_g\in(0.6, 6)$ and $z_{\rm tail}\in(0.1, 5)$. The resulting constraints are shown in Fig. \ref{fig:ztail_nz} for the constant-amplitude (orange contours) and constant-bias (black-gray contours) redshift dependent models of $b_g(z)$. Looking at the constant-amplitude constraints, the uncertainty on $b_g$ ($\sigma(b_g)=0.28$) degrades significantly with respect to the previous results where the redshift distribution was fixed to the SKADS estimate. The high-redshift tail, on the other hand, is constrained to be $z_{\rm tail}=1.30^{+0.27}_{-0.40}$. In the constant-bias case, a larger value of $b_g$ is preferred, to compensate for the lack of growth as a function of redshift, which is accompanied by a preference for larger values of $z_{\rm tail}$ and overall larger uncertainties on both parameters. In both cases, however, the data show a preference for larger redshift tails than that implied by the photometric redshifts included in the LoTSS value-added catalog. This can be seen explicitly in Figure \ref{fig:nz_tail}, which shows the 1$\sigma$ bounds on the recovered redshift distribution in comparison with the estimates from the LoTSS VAC, VLA-COSMOS and SKADS.

      The corresponding constraints on $b_g(z)$ are shown in Fig. \ref{fig:bzs} in both cases. The true redshift evolution of the effective galaxy bias for the LoTSS sample analysed here is most likely not constant but potentially less steep than the constant-amplitude model, therefore lying somewhere between these two extremes. To illustrate this, the figure also shows the bias values measured for different radio populations by \citet{Hale:2017wub}, and the NVSS sample of \citet{2015ApJ...812...85N}. The results above therefore imply that the true underlying distribution is probably more compatible with the SKADS or VLA-COSMOS estimates than the distribution of photometric redshifts in the LoTSS VAC.

    \subsection{Constraints on bias and $\sigma_8$}\label{ssec:results.bias_s8}
      Under the assumption that the true underlying redshift distribution of the LoTSS sample is well described by the SKADS estimate, we can use the combination of $\hat{C}_\ell^{gg}$ and $\hat{C}_\ell^{g\kappa}$ to break the degeneracy between $b_g$ and the amplitude of matter fluctuations parametrized by $\sigma_8$. Given the existing redshift distribution uncertainties, and the degeneracy with other cosmological parameters, the resulting constraints would be neither robust nor competitive. The aim of this exercise is therefore twofold: a sanity check to verify that our data is broadly compatible with the standard cosmological model, and a demonstration of the use of continuum surveys for cosmology.

\begin{figure}
        \begin{center}
          \includegraphics[width=0.49\textwidth]{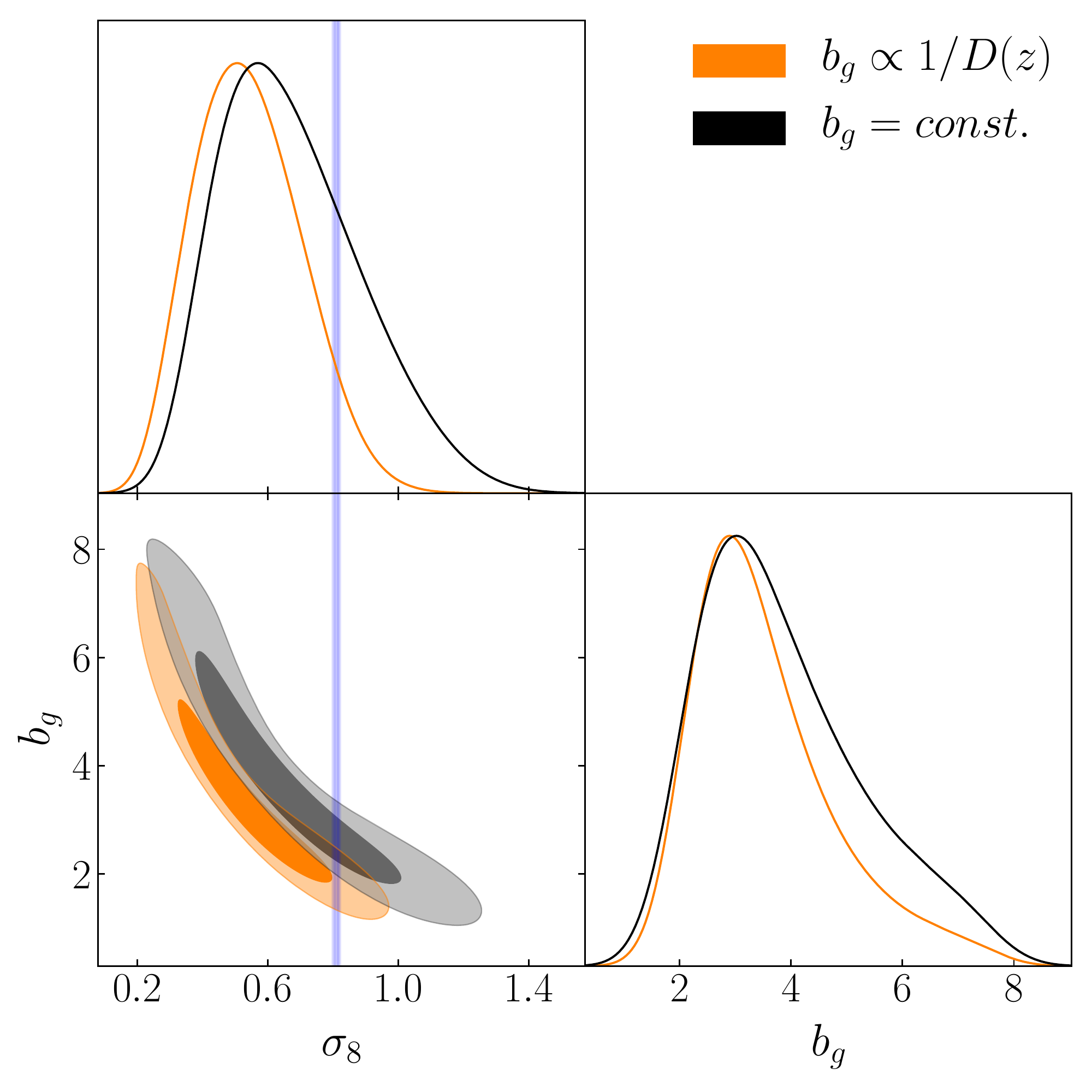}
          \caption{Constraints on the galaxy bias  parameter $b_g$ and the amplitude of matter inhomogeneities $\sigma_8$ from the combination of $C_\ell^{gg}$ and $C_\ell^{g\kappa}$. Results are shown for a constant-bias model (black contours) and for a bias that evolves with the inverse of the linear growth factor (orange contours). The blue band shows the constraints on $\sigma_8$ found by \planck{} \citet{2018arXiv180706209P}. Our power spectrum measurements are in good agreement with the standard $\Lambda$CDM model as constrained by \planck{}.}\label{fig:bs8}
        \end{center}
      \end{figure}

      The results are shown in Figure \ref{fig:bs8} for the constant-bias and constant-amplitude redshift dependent models. The constraints on $\sigma_8$ for both bias models are
      \begin{align}\nonumber
        &\sigma_8=0.69^{+0.14}_{-0.21},\hspace{6pt}{\rm (constant\,\,amplitude)},\\
        &\sigma_8=0.79^{+0.17}_{-0.32},\hspace{6pt}{\rm (constant\,\,bias)}.
      \end{align}
      Both values are in agreement with each other as well as with those those found by \planck{} ($\sigma_8=0.811\pm0.006$, shown in blue in the figure), with a preference for lower $\sigma_8$ values along the $b_g$-$\sigma_8$ degeneracy direction.

  \section{Conclusions}\label{sec:conc}
    Radio continuum surveys are a valuable probe to study the properties and evolution of star-forming galaxies and AGNs. Unimpeded by dust attenuation, radio surveys cover a much broader range of redshifts, and correspondingly larger volumes, than their optical counterparts. Because of this, continuum surveys also constitute a potential avenue to reconstruct the growth of structure over the largest scales, and therefore have drawn the interest of the cosmology community.

    However, the lack of redshift information from radio continuum data makes these surveys reliant on matching to deep optical catalogs to constrain the radial distribution of the sources, and to reconstruct the redshift evolution of their properties. The potential incompleteness of the optical cross-matches makes redshift calibration one of the largest sources of systematic uncertainty in the potential use of radio continuum surveys for cosmology. This is akin to the challenges faced by ongoing and future photometric weak lensing surveys when characterising and propagating the uncertainties in the redshift distribution of galaxies \citep{2020arXiv200409542S,2020arXiv200715635H,2020arXiv200712795S}.

    In this paper we have studied the clustering of galaxies on the flux-density limited \lotss{} first data release, both through the harmonic-space auto-correlation, and through their cross-correlation with the lensing convergence of the CMB, as measured by \planck{}. The cross-correlation is detected at the $5\sigma$ level.

    To illustrate the challenge posed by the uncertainties in the redshift distribution $dp/dz$, we have considered three estimates of this quantity: from the LoTSS value-added catalog, from the VLA-COSMOS cross-matched sample (which includes both spectroscopic and photometric redshifts), and from the SKADS simulations. We have shown that the truncated high-redshift tail from the LoTSS-VAC photometric redshifts leads to radically different interpretations of the clustering amplitude when compared with the results from the VLA-COSMOS or SKADS redshift distributions. On the other hand, the CMB lensing cross-correlation alone is fairly insensitive to variations in $dp/dz$, and leads to consistent measurements of the galaxy bias.

    The robustness of the CMB-lensing cross-correlation to redshift distribution uncertainties, allows us to break the degeneracy between the galaxy bias and the width of the $dp/dz$ by combining it with the clustering auto-correlation. Through this joint analysis, we are able place constraints on the high-redshift tail of the distribution, showing that it is underestimated by the LoTSS-VAC photo-$z$s, and better represented by the VLA-COSMOS and SKADS estimates. To our knowledge, this is the first attempt in the literature at calibrating redshift distributions through cross-correlations with CMB lensing. This is akin to the ``cross-correlation redshifts'' approach, based on using cross-correlations against samples with a known redshift distribution \citep{2008ApJ...684...88N,2017PhRvD..96d3515A,2018MNRAS.477.1664G} (in this case, the CMB lensing kernel). Given the broad range of redshifts covered by the CMB lensing kernel, it is unlikely that this approach will be able to constrain a shift in the mean redshift of the target distribution \citep[e.g. see][]{2017PhRvD..96d3515A}, as is usually done in weak lensing analyses \citep{2020arXiv200715635H}, but it should be useful to calibrate the width of the distribution or the presence of photo-$z$ outliers. This could provide a robust way to extract cosmological information from samples with poor spectroscopic coverage in future surveys, such as the Legacy Survey of Space and Time \citep[LSST;][]{2018arXiv180901669T}.

    Finally, assuming that the redshift distribution is well described by the SKADS estimate (which is broadly consistent with the VLA-COSMOS redshift distribution), we use the combination of clustering and CMB lensing to measure the amplitude of matter inhomogeneities $\sigma_8$. Although the resulting constraints are neither competitive nor robust, given the existing uncertainties on $dp/dz$ and on the redshift dependence of the galaxy bias, this exercise allows us to demonstrate the use of continuum surveys for cosmology. The measured value of $\sigma_8=0.79^{+0.17}_{-0.32}$, assuming a constant bias, agrees well with the constraints found by CMB and large-scale structure experiments \citep{2018arXiv180706209P,2018PhRvD..98d3526A,2019PASJ...71...43H,2020arXiv200715633A}.
    
    This paper highlights the novel science that can be carried out by combining CMB and radio-continuum survey using the first data release from \lotss{}, which covers 424 square degrees. \lotss{} itself should cover the vast majority of the northern sky when completed ($\sim 20,000$ square degrees, allowing much stronger constraints on the various results presented here. Furthermore, the EMU Survey \citep{Norris2011} will provide a complementary southern hemisphere survey, down to a similar equivalent flux-density limit at GHz frequencies. However, as we have shown in this work, one of the key uncertainties in using radio continuum surveys for cosmology arises from the lack of a well constrained redshift distribution. Fortunately, there are a wealth of surveys that will aid in pinning this down on the timescale of these all-sky surveys. Future \lotss{} data releases will be able to make use of deep fields with $\sim90\%$ photometric redshift coverage. Furthermore, the WEAVE-LOFAR Survey \citep{Smith2016} will provide a huge number of spectroscopic redshifts for the \lotss{} sources, combining deep spectroscopy in the LOFAR deep survey fields, with spectroscopy of the rarer brighter sources, through a tiered radio-continuum selected survey. However, spectroscopic surveys are rarely complete, and tend to be limited by the optical magnitude depth or the ability to detect emission lines (an advantage for radio continuum selected surveys). As such, further important information will be gleaned from measuring the redshift distribution of radio sources in the deep extragalactic survey fields, similar to how we have used the VLA-COSMOS survey here. However, a single deep field is inherently limited by sample variance, as such the MeerKAT International Giga-Hertz Tiered Extragalactic Exploration  \citep[MIGHTEE; ][]{Jarvis2016} Survey, which will cover $\sim 20$ square degrees over four extragalactic deep fields, accessible from the southern hemisphere, will help pin down the redshift distribution of radio continuum sources to very deep flux-density limits using both photometric \citep[e.g.][]{Jarvis2013, Adams2020} and spectroscopic \citep[e.g.][]{2018MNRAS.480..768D,2014SPIE.9147E..0NC,2019Msngr.175...46D} redshifts.
    
    Taken together these advances will allow us to begin to divide the radio continuum source population into distinct sub-samples, e.g. SFGs and AGN. This would then allow us to introduce more realistic bias evolution models for these different populations, with different redshift distributions, which is a limitation on our current work due to the relatively small survey area of the \lotss{} DR1. 
    
    Looking further in the future, the sky surveys planned for the Square Kilometre Array will lead to higher number density of objects across huge swathes of sky. Although this may not substantially increase the number of AGN at high redshift ($z>2$), these surveys  will increase the number of SFGs across all redshifts, and allow the possibility of a more robust decoupling of the SFG and AGN populations \citep[e.g.][]{2015aska.confE..81M,2020MNRAS.495.1188M}, given the much higher angular resolution ($<1$\,arcsec). This could also be achieved by LOFAR using its international baselines.

  \section*{Data availability statement}
    The data underlying this article are available in the \planck{} legacy archive \url{https://www.cosmos.esa.int/web/planck/pla}, the LOFAR surveys website \url{https://lofar-surveys.org/releases.html}, and the VLA-COSMOS 3 GHz repository at the Strasbourg astronomical Data Center \url{http://cdsarc.u-strasbg.fr/viz-bin/qcat?J/A+A/602/A2}.

  \section*{Acknowledgements}
    We would like to thank Emmanuel Schaan for useful comments and discussions. DA acknowledges support from the Beecroft Trust, and from the Science and Technology Facilities Council through an Ernest Rutherford Fellowship, grant reference ST/P004474. EB is supported by the European Research Council Grant No: 693024  and  the  Beecroft  Trust. MJJ acknowledges support from the UK Science and Technology Facilities Council [ST/N000919/1], the South African Radio Astronomy Observatory (SARAO; www.ska.ac.za) and the Oxford Hintze Centre for Astrophysical Surveys which is funded through generous support from the Hintze Family Charitable Foundation. DJS acknowledges the Research Training Group 1620 ‘Models of Gravity’, supported by Deutsche Forschungsgemeinschaft (DFG) and support by the German Federal Ministry for Science and Research BMBF-Verbundforschungsprojekt D-LOFAR IV (grant number 05A17PBA). LOFAR data products were provided by the LOFAR Surveys Key Science project (LSKSP\footnote{\url{https://lofar-surveys.org/}}) and were derived from observations with the International LOFAR Telescope (ILT). LOFAR \citep{vanHaarlem2013} is the Low Frequency Array designed and constructed by ASTRON. It has observing, data processing, and data storage facilities in several countries, that are owned by various parties (each with their own funding sources), and that are collectively operated by the ILT foundation under a joint scientific policy. The efforts of the LSKSP have benefited from funding from the European Research Council, NOVA, NWO, CNRS-INSU, the SURF Co-operative, the UK Science and Technology Funding Council and the Jülich Supercomputing Centre. This work uses data based on observations obtained with Planck (\url{http://www.esa.int/Planck}), an ESA science mission with instruments and contributions directly funded by ESA Member States, NASA, and Canada. Contour plots were generated using the {\tt GetDist} package \citep{Lewis:2019xzd}. Some of the results in this paper have been derived using the HEALPix \citep{2005ApJ...622..759G} package. We made extensive use of the {\tt scipy} \citep{2020SciPy-NMeth}, {\tt healpy} \citep{Zonca2019}, and {\tt matplotlib} \citep{Hunter:2007} python packages.

\setlength{\bibhang}{2.0em}
\setlength\labelwidth{0.0em}
\bibliography{main}

\end{document}